\documentclass[12pt]{l4dc2021} 
\title[Primal-dual Learning for the Model-free Risk-constrained LQR]{Primal-dual Learning for the Model-free Risk-constrained Linear Quadratic Regulator}
\usepackage{times}
\usepackage{appendix}
\allowdisplaybreaks[1]
\newtheorem{assum}{Assumption}

\author{\Name{Feiran Zhao} \Email{zhaofr18@mails.tsinghua.edu.cn}\\
	\Name{Keyou You} \Email{youky@tsinghua.edu.cn}\\
	\addr Department of Automation and BNRist, Tsinghua University}
\begin{document}	
	\maketitle	
	\begin{abstract}
	Risk-aware control, though with promise to tackle unexpected events, requires a known exact dynamical model. In this work, we propose a model-free framework to learn a risk-aware controller of a linear system. We formulate it as a discrete-time infinite-horizon LQR problem with a state predictive variance constraint. Since its optimal policy is known as an affine feedback, i.e., $u^*(x) = -Kx+l$, we alternatively optimize the gain pair $(K,l)$ by designing a primal-dual learning algorithm. First, we observe that the Lagrangian function enjoys an important local gradient dominance property. Based on it, we then show that there is no duality gap despite the non-convex optimization landscape. Furthermore, we propose a primal-dual algorithm with global convergence to learn the optimal policy-multiplier pair. Finally, we validate our results via simulations.
	\end{abstract}	

\section{Introduction}

Stochastic optimal control~\citep{aastrom2012introduction} is a well-studied framework that deals with inherent random noises in the dynamical system. Its classical formulation targets to minimize an expected long-term cost, which is risk-neutral as it
only optimize the expectation without explicit considerations on the variability of the state. Thus, the system behaviours may be easily influenced by less probable but large noises, leading to catastrophic consequences for the safety-critical systems. In decades, the risk-aware controllers have been proposed to tackle the extreme noises with a slight sacrifice of average performance~\citep{sopasakis2019risk, jacobson1973optimal,moore1997risk, bauerle2014more,roulet2020convergence}. For example, the risk is typically addressed by replacing the cost with its exponentiation~\citep{speyer1992optimal,moore1997risk} or optimizing the risk measure~\citep{chapman2019risk} e.g., Conditional Value-at-Risk (CVaR)~\citep{rockafellar2000optimization}. However, most of them are model-based~\citep{chapman2019risk, moore1997risk, speyer1992optimal} and hence not directly applicable when the exact dynamical model is unknown.

Model-free reinforcement learning (RL)~\citep{sutton1998introduction, bertsekas2019reinforcement} has achieved tremendous progress recently in the continuous control field~\citep{mnih2015human-level,lillicrap2016continuous}. Instead of identifying the underlying dynamical model first, it approaches the control problem by directly searching for an optimal policy that minimizes the estimated cost function. Under the RL framework, the prevalent risk-averse methods~\citep{wen2018constrained, prashanth2018risk, borkar2014risk} take the risk into consideration by, e.g., adding a risk-related cumulative cost constraint to the Markov decision process (MDP)~\citep{paternain2019constrained, chow2017risk, yu2019convergent, tessler2018reward}, or formulating the risk as an adversary~\citep{pan2019risk}. Though empirically successful on the continuous control benchmarks~\citep{pan2019risk, tessler2018reward}, they typically lack strong theoretical guarantees, hampering their physical-world applications. 

Recent advances in the context of policy optimization (PO) for the linear quadratic regulator (LQR)~\citep{bertsekas1995dynamic}, including policy gradient~\citep{fazel2018global,bu2019lqr,zhang2019policy,zhang2020policy} and random search methods~\citep{malik2019derivative,mohammadi2020linear}, have been shown to enjoy the global convergence in spite of the non-convex nature of the optimization landscape. Some works focus on the LQR variants e.g., robust control with multiplicative noises~\citep{gravell2020learn}, distributed LQR~\citep{li2019distributed} and Markov jump linear systems\citep{jansch2020convergence}. In particular, the PO for $\mathcal{H}_2$ linear control with $\mathcal{H_{\infty}}$ robustness guarantees is analyzed in~\citet{zhang2020policy} for a risk-sensitive linear exponential quadratic Gaussian (LEQG)~\citep{whittle1981risk} instance. However, to the best of our knowledge, there is no such analysis for the risk-aware formulation with a risk constraint explicitly concerned.

%Although the risk-sensitive formulation is studied in, it requires the noise to be Gaussian with zero-mean. 
%In particular, there is no global convergence guarantees for such algorithms. Moreover, the stability of the resultant closed-loop system is not concerned, which is the fundamental requirement in control theory. Alongside, we take an initial step towards understanding the theoretical aspects of PO for the constrained LQR.

In this paper, we consider the learning problem for the model-free risk-aware controller. Inspired by~\citet{tsiamis2020risk}, we formulate it as a discrete-time infinite-horizon LQR problem with a one-step predicted state variability constraint. By~\citet{zhao2021infinitehorizon}, the solution to it is an affine state feedback policy. Thus, we can alternatively optimize over the stabilizing affine policy set. Nevertheless, in contrast to LQR, three challenges exist in our setting. Firstly, the constraint optimization problem is non-convex in that the objective function, the risk constraint and the stabilizing policy set are all non-convex. Moreover, the optimization variable in LQR is a single feedback gain~\citep{fazel2018global}, while in our case it is a gain pair and hence the optimization landscape is not clear yet. Finally, the first-order optimization methods cannot be used since the dynamical model is unknown. 

This work proposes a primal-dual learning framework to solve the risk-constrained LQR problem. Alongside, we take an initial step towards understanding the theoretical aspects of PO for the constrained LQR. Our contributions are summarized below. Firstly, in spite of the constrained non-convex optimization nature, we show that the strong duality holds. Secondly, we study the optimization landscape of the Lagrangian function over the stabilizing affine policy set. In particular, we find that it enjoys two favourable properties, i.e., the local gradient dominance and Lipschitz property. Thirdly, we propose a primal-dual algorithm to learn the optimal policy-multiplier pair and show its global convergence.

\section{Problem Formulation}
In the standard setup of LQR, we consider a time-invariant discrete linear stochastic system with full state observations,
\begin{equation}\label{equ:sys}
x_{t+1} = Ax_t + Bu_t+w_t,
\end{equation}
where the next state $x_{t+1}$ is a linear combination of the current state $x_t \in \mathbb{R}^n$, the control $u_t \in \mathbb{R}^m$, and the random noise $w_t\in \mathbb{R}^d$. The model parameters are denoted as $A\in \mathbb{R}^{n \times n}$ and $B \in \mathbb{R}^{n \times m}$. 

The goal of infinite-horizon LQR is to find a control policy $\pi$ which minimizes an average long-term cost, i.e.,
\begin{equation}\label{prob:lqr}
\begin{aligned}
\text { minimize } &~~ \lim\limits_{T\rightarrow \infty} \frac{1}{T} \mathbb{E}  \sum_{t=0}^{T-1}(x_{t}^{\top} Q x_{t}+u_{t}^{\top} R u_{t})\\
\end{aligned}
\end{equation}
where $u_t = \pi(h_t)$ with the history trajectory $h_t = \{x_0,u_0,\cdots, x_{t-1},u_{t-1}\}$ and the expectation is taken with respect to the random noise $w_t$. Throughout the paper, we make the following assumption standard in the control theory~\citep{bertsekas1995dynamic}.
\begin{assum}
	\label{assumption}
	$Q$ is positive semi-definite and $R$ is positive definite. The pair $(A,B)$ is stabilizable and $(A,Q^{{1}/{2}})$ is observable.
\end{assum}

Under Assumption \ref{assumption}, solving (\ref{prob:lqr}) yields a unique linear state feedback policy $u_t = -Kx_t$ when $w_t$ has zero mean. Clearly, the classical LQR is risk-neutral as it aims to minimize only the expected cost. Thus, the state may be largely influenced by the low-probability but large noises, especially those with heavy-tailed distributions.

In this paper, we study the infinite-horizon risk-constrained LQR in~\cite{zhao2021infinitehorizon} and solve it in a model-free approach. That is,
\begin{equation}\label{equ:rclqr}
\begin{aligned}
\text { minimize } &~~ \lim\limits_{T\rightarrow \infty} \frac{1}{T} \mathbb{E}  \sum_{t=0}^{T-1}(x_{t}^{\top} Q x_{t}+u_{t}^{\top} R u_{t})\\
\text { subject to }& ~~\lim\limits_{T\rightarrow \infty} \frac{1}{T}  \mathbb{E} \sum_{t=0}^{T-1}(x_{t}^{\top} Q x_{t}-\mathbb{E}[x_{t}^{\top} Q x_{t}|h_t])^{2} \leq \rho \\
\end{aligned}
\end{equation}
where $\rho > 0$ is a user-defined risk tolerance constant. In contrast to standard LQR (\ref{prob:lqr}), we do not require the noise $w_t$ to be zero-mean. Instead, we only assume a finite 4th-order moment of $w_t$~\citep{tsiamis2020risk}.

In our recent work~\cite{zhao2021infinitehorizon}, we have shown that the optimal policy to (\ref{equ:rclqr}) is an affine state feedback, i.e., $u^*(x) = -K^*x +l^*$, which is also able to stabilize the system. Exploiting this affine structure, we can alternatively optimize the gain pair $(K,l)$. Define the mean $\bar{w} = \mathbb{E}[w_t]$, the covariance $W = \mathbb{E}[(w_t - \bar{w})(w_t - \bar{w})^{\top}]>0$, higher-order weighted statistics 
$
M_{3} = \mathbb{E}[(w_{i}-\bar{w})(w_{i}-\bar{w})^{\top} Q(w_{i}-\bar{w})] \text { and } 
m_{4} = \mathbb{E}[(w_{i}-\bar{w})^{\top} Q(w_{i}-\bar{w})-\operatorname{tr}(W Q)]^{2} 
$
of the noise $w_t$. Given that $w_t$ has a finite 4-order moment, (\ref{equ:rclqr}) can be reformulated by~\cite{zhao2021infinitehorizon} as
\begin{equation}\label{prob:new_rclqr}
\begin{aligned}
\text { minimize } &~~ J(K,l) = \lim\limits_{T\rightarrow \infty} \frac{1}{T} \mathbb{E}  \sum_{t=0}^{T-1}(x_{t}^{\top} Q x_{t}+u_t^{\top} R u_t)\\
\text { subject to }& ~~ J_c(K,l) = \lim\limits_{T\rightarrow \infty} \frac{1}{T}  \mathbb{E} \sum_{t=0}^{T-1}(4x_{t}^{\top} QWQ x_{t} + 4x_{t}^{\top}QM_3  ) \leq \bar{\rho} \\
\end{aligned}
\end{equation}
with $u_t = -Kx_t + l$ and $\bar{\rho} = \rho - m_4 + 4\operatorname{tr}\{(WQ)^2\}$. In (\ref{prob:new_rclqr}), $(K,l)$ is the optimization variable.

The direct PO for the risk-neutral formulation (\ref{prob:lqr}) has been well studied and typically enjoys the convergence guarantee, including random search and policy gradient methods~\cite{fazel2018global}. However, (\ref{prob:new_rclqr}) is not only a non-convex constrained optimization problem, but also differs from (\ref{prob:lqr}) in that the optimization variable is a gain pair $(K,l)$. In this paper, we study the analytical property of the constrained optimization problem (\ref{prob:new_rclqr}). Furthermore, we propose a convergent primal-dual algorithm to solve it exactly by solely using data.

\section{Primal-dual Optimization for Risk-constrained LQR}\label{sec:primal-dual}
In this section, we introduce the primal-dual method for solving the risk-constrained LQR problem in (\ref{prob:new_rclqr}). In contrast to \cite{fazel2018global}, its Lagrangian function is only locally gradient dominated and locally Lipschitz with respect to the policy. Moreover, we establish the strong duality for the non-convex constrained optimization problem (\ref{prob:new_rclqr}).

\subsection{Primal-dual method}
In the rest of the paper, we use the augmented matrix $X = [K~l]$ to denote the optimization variable. Define $\mathcal{S} = \{X=[K ~l] | \rho(A-BK) <1, K \in \mathbb{R}^{n \times m}, l \in \mathbb{R}^{n} \}$, where $\rho(\cdot)$ denotes the spectral radius. Clearly, we have $J(X) < +\infty$ and $J_c(X)< +\infty$ if and only if $X \in \mathcal{S}$. Let $\mu \geq 0$ denote the Lagrange multiplier and $Q_{\mu} = Q+4 \mu Q W Q$ and $S = 2\mu QM_3$. We define the Lagrangian function of (\ref{prob:new_rclqr}) as
\begin{equation}\label{def:L}
\mathcal{L}(X,\mu) = J(X)+\mu(J_c(X) - \bar{\rho})
= \lim\limits_{T\rightarrow \infty} \frac{1}{T} \mathbb{E}  \sum_{t=0}^{T-1}c_\mu(x_t, u_t),
\end{equation}
where $c_\mu(x_t, u_t)=x_{t}^{\top} Q_{\mu} x_{t}+ 2x_{t}^{\top}S +u_{t}^{\top} R u_{t}- \mu \bar{\rho}$, which is a reshaped cost with a risk weight $\mu$ that balances the objective and the risk. Accordingly, we define the dual function
$
D(\mu) = \min_{X} \mathcal{L}(X,\mu)
$
and the dual problem
\begin{equation}\label{prob:dual}
\max \limits_{\mu \geq 0}~ D(\mu) = \max \limits_{\mu \geq 0}~ \min _{X \in \mathcal{S}} \mathcal{L}(X, \mu).
\end{equation}

Our primal-dual method is iteratively given as
\begin{align}
X_j&\in \mathop{\text{argmin}} \limits_{X \in \mathcal{S}}~ \mathcal{L}(X,\mu_j), \label{prim_iterate} \\
\mu_{j+1}&= [\mu_j+ \xi_j \cdot \omega(\mu_j)]_+, \label{dual_iterate}
\end{align}
where the stepsize $\xi_j>0$, $\omega(\mu_j)$ is a subgradient of $D(\mu)$ at $\mu_j$ and $[x]_+=\max\{0, x\}$ for any $x\in\mathbb{R}$.

To guarantee the global convergence of the primal-dual method,  the strong duality between the primal problem and dual problem is essential.  However, the constrained optimization problem (\ref{prob:new_rclqr}) is non-convex and, therefore, the strong duality does not trivially follow. Moreover, the primal-dual method requires to solve (\ref{prim_iterate}) under a fixed multiplier $\mu$. Though for LQR problems some model-free algorithms are guaranteed to find an optimal gain $K$~\citep{fazel2018global, malik2019derivative,mohammadi2020linear}, they cannot be directly applied as our optimization variable is a gain pair $(K,l)$. In particular, these algorithms exploit favourable properties of the objective function such as gradient dominance~\citep{fazel2018global} and Lipschitz continuity~\citep{malik2019derivative}, which are unclear for $\mathcal{L}(K,l,\mu)$. In what follows, we work towards addressing these problems.

\subsection{Closed-form of the Lagrangian Function and its Gradient}\label{subsec:closed-form}

We first derive the closed-form of $\mathcal{L}(X,\mu)$. It follows from (\ref{def:L}) that $\mathcal{L}(X,\mu)$ is finite if and only if $X \in \mathcal{S}$. For a stabilizing policy $X \in \mathcal{S}$, the state has a stationary distribution, the mean $\bar{x}_{K,l}$ of which satisfies 
$\bar{x}_{K,l} = (A-BK)\bar{x}_{K,l} + Bl + \bar{w}$,
and its correlation matrix can be solved through a Lyapunov equation
\begin{equation}
\Sigma_{K}=W+(A-B K) \Sigma_{K}(A-B K)^{\top}.
\end{equation}
Suppose that $P_K \geq 0$ is the solution of the Lyapunov equation
$$
P_{K} = Q_{\mu} +K^{\top} R K + (A-B K)^{\top} P_{K}(A-B K)
$$
and let $E_{K}=(R+B^{\top} P_{K} B) K-B^{\top} P_{K} A$ and $V = (I - (A-BK))^{-1}$. 
\begin{proposition}[Closed-form expression]
	The Lagrangian function $\mathcal{L}(X,\mu)$ is given by
	\begin{equation}\label{equ:closed-form}
	\mathcal{L}(K,l,\mu) = \mathrm{tr}\{P_K(W + (Bl+\bar{w})(Bl+\bar{w})^{\top})\} + g_{K,l}^{\top}(Bl+\bar{w})  + l^{\top}Rl - \mu \bar{\rho}.
	\end{equation}
	where $g_{K,l}^{\top} = 2(-l^{\top}E_K+S^{\top} + \bar{w}^{\top}P_K(A-BK))V$ and $z_{K,l}$ is a constant.
\end{proposition}

\begin{proposition}[Policy gradient expression]\label{prop:grad}
	The gradient of $\mathcal{L}(X,\mu)$ with respect to $X$ is given by
	$
	\nabla \mathcal{L}(X,\mu) = 2\left[
	E_K~~
	G_{K,l}
	\right] \Phi_{K,l},
	$
	where $G_{K,l} = (R+B^{\top}P_KB)l + B^{\top}P_K\bar{w} + \frac{1}{2} B^{\top}g_{K,l}$ and $\Phi_{K,l}$ is the correlation matrix
	\begin{equation}\label{def:phi}
	\Phi_{K,l} = \lim\limits_{T\rightarrow \infty} \frac{1}{T}\mathbb{E}\sum_{t=0}^{T-1} 		
	\begin{bmatrix}
	x_t \\
	-1
	\end{bmatrix}
	\begin{bmatrix}
	x_t \\
	-1
	\end{bmatrix}^{\top}=
	\begin{bmatrix}
	\Sigma_K + \bar{x}_{K,l}\bar{x}_{K,l}^{\top} &  -\bar{x}_{K,l}\\
	-\bar{x}_{K,l}^{\top} & 1
	\end{bmatrix}>0.
	\end{equation}
\end{proposition}

Since $\Phi_{K,l}$ is positive definite, the stationary point of $\mathcal{L}(X, \mu)$ can be uniquely solved by setting the gradients to zero as $X^*(\mu) = [K^*(\mu)~ l^*(\mu)] $ with
\begin{equation}\label{equ:stationary}
\begin{aligned}
&K^*(\mu) = (R+B^{\top} P_{K^*(\mu)} B)^{-1}B^{\top} P_{K^*(\mu)} A, \\
&l^*(\mu) = -(R+B^{\top}P_{K^*(\mu)}B)^{-1}B^{\top}V^{\top}(P_{K^*(\mu)}\bar{w}+S). \\
\end{aligned}
\end{equation}

%We remark that in contrast to the classical LQR, the steady state has a deviation from the origin due to the weighted risk cost in $c_\mu(x_t, u_t)$. This can be observed from the linear term $g_{K,l}^{\top}x$ in the value function (\ref{equ:value}). The expression of $\mathcal{L}(K,l,\mu)$ is further complicated by the non-zero mean $\bar{w}$ and $l$, which are both zero in LQR. Fortunately, its gradient can be written in a simple form by reorganizing (\ref{equ:gradient}). When $\bar{x}_{K,l} = 0$, $\nabla_{K} \mathcal{L}(K,l,\mu)$ naturally reduces to its LQR counterpart.

\subsection{Properties of the Lagrangian Function}
The minimization on $\mathcal{L}(X,\mu)$ in (\ref{equ:closed-form}) is a non-convex optimization problem, in that both the objective function and the stabilizing policy set $\mathcal{S}$ are non-convex, which poses challenges in solving (\ref{prim_iterate}) with standard policy gradient-based methods. In the PO for classical LQR problems (\ref{prob:lqr})~\citep{fazel2018global}, this is alleviated by observing that the objective function is globally gradient dominated. For a differentiable function $f(x): \mathbb{R}^n \rightarrow \mathbb{R}$ with a finite global minimum $f^*$, it is {\em globally gradient dominated} if
\begin{equation}\label{dominated}
f(x) - f^* \leq \lambda \| \nabla f(x) \|^2, ~~\forall x \in \text{dom}(f)\subseteq \mathbb{R}^n
\end{equation}
where $\lambda \geq 0$ is a {gradient dominance constant}. Clearly, it implies that a stationary point must be the global minimizer. Hence, if $f(z)$ is also Lipschitz smooth, one would expect that the gradient-based algorithms converge at a linear rate to the global minimum~\citep{malik2019derivative}. 

We show that $\mathcal{L}(X,\mu)$ enjoys a local gradient dominance property, which is weaker than the more common global one in the sense that it only holds locally over a compact set. Before formalizing it, we note that the compact set can be constructed by observing that $\mathcal{L}(X,\mu)$ is coercive. 
\begin{lemma}[Coercivity]Under a fixed $\mu>0$, the Lagrangian
	$\mathcal{L}(X,\mu)$ is coercive in $X$ in the sense that
	$
	\lim \limits_{X \rightarrow \partial \mathcal{S}} \mathcal{L}(X,\mu) = +\infty,
	$
	where $\partial \mathcal{S}$ denotes the boundary of $\mathcal{S}$. Moreover, it has a compact $\alpha$-sublevel set
	\begin{equation}\label{def:sublevel}
	\mathcal{S}_{\alpha}  = \{ X \in \mathbb{R}^{m\times {(n+1)}}|\mathcal{L}(X,\mu) \leq \alpha \}.
	\end{equation}
\end{lemma}

Then, we obtain the local gradient dominance property of $\mathcal{L}(X, \mu)$ over $\mathcal{S}_{\alpha}$.
\begin{lemma}[Local Gradient Dominance]\label{lem:gradient dominance}
	$\mathcal{L}(X, \mu)$ is gradient dominated locally over the compact set $\mathcal{S}_{\alpha}$ in (\ref{def:sublevel}), namely,
	\begin{equation}\notag
	\mathcal{L}(X,\mu)-\mathcal{L}(X^*(\mu),\mu)
	\leq \lambda_{\alpha}\operatorname{tr} \{
	\nabla\mathcal{L}^{\top}\nabla\mathcal{L}	
	\},
	\end{equation}
	where $\lambda_{\alpha} = \frac{\|\Phi^*\|}{4\sigma_{min}(R) \cdot \phi_{\alpha}^2} >0$ is a constant related to $\mathcal{S}_\alpha$ and $\phi_{\alpha} = \min \limits_{X \in \mathcal{S}_{\alpha}}  \sigma_{min}(\Phi_{K,l})>0$.
\end{lemma}

By the local gradient dominance and the coercivity, we can determine the global minimizer of the Lagrangian.
\begin{theorem}\label{theorem:unique}
	The critial point $X^*(\mu)$ in (\ref{equ:stationary}) is the unique global minimizer of $\mathcal{L}(X,\mu)$.
\end{theorem}

Finally, we show that both $\mathcal{L}(X,\mu)$ and its gradient $\nabla \mathcal{L}$ are locally Lipschitz.
\begin{lemma}[Locally Lipschitz]\label{lem:lip}
	There exist positive scalars $(\zeta_{X}, \beta_X, \gamma_X)$ that depends on the current policy $X$, such that for all policies $X' \in \mathcal{S}$ satisfying $\| X'-X\| \leq \gamma_X$, we have
	\begin{align*}
	&|\mathcal{L}(X',\mu) - \mathcal{L}(X,\mu)| \leq \zeta_X \| X'-X\|  ~~\text{and}~\\
	&\|\nabla \mathcal{L}(X',\mu) - \nabla \mathcal{L}(X,\mu)\| \leq \beta_X \| X'-X\|.
	\end{align*}
\end{lemma}
Note that the scalars $\zeta_{X}, \beta_X, \gamma_X$ in Lemma \ref{lem:lip} are functions of $X$ as well as the problem parameters e.g., $(A,B,Q,R)$.

\subsection{Strong Duality}
The duality analysis is generally difficult for a non-convex constrained optimization problem. Nevertheless, we show that the strong duality between the primal problem  (\ref{prob:new_rclqr}) and dual problem \eqref{prob:dual} indeed holds, by leveraging the established properties of the Lagrangian.

\begin{theorem}\label{theorem:duality}
	Suppose that the Slater's condition in \eqref{prob:new_rclqr} holds, i.e., there exists a policy $\widetilde{X} \in \mathcal{S}$ such that $J_c(\widetilde{X})<\bar{\rho}$, then there is no duality gap for the primal problem (\ref{prob:new_rclqr}) and the dual problem \eqref{prob:dual}.
\end{theorem}

\section{Primal-dual Learning Algorithm for the Risk-constrained LQR}

In the model-free setting, $(A,B)$ is unknown and the gradient $\nabla \mathcal{L}(X,\mu)$ cannot be computed directly. Thus, we estimate the gradient via noisy samples of the Lagrangian. By focusing on a sublevel set, we can leverage the gradient dominance and smoothness to develop a random search method to solve (\ref{prim_iterate}). Moreover, we propose a primal-dual algorithm to find an optimal pair $(X^*, \mu^*)$ where an estimation of the subgradient is also used for the dual ascent in (\ref{dual_iterate}).

\subsection{Random Search for (\ref{prim_iterate})}
Assume that we have a cost oracle, which returns a noisy evaluation of $\mathcal{L}(X,\mu)$ and $J_c(X)$ as
%\begin{align*}
%	\widehat{\mathcal{L}}(X,\mu)& =  \lim\limits_{T\rightarrow \infty} \frac{1}{T} \sum_{t=0}^{T-1}c_\mu(x_t, u_t) - \mu \bar{\rho} ~~~~ \text{and} \\
%	\widehat{J}_c(X) & = \lim\limits_{T\rightarrow \infty} \frac{1}{T}  \sum_{t=0}^{T-1}(4x_{t}^{\top} QWQ x_{t} + 4x_{t}^{\top}QM_3),
%\end{align*}
\begin{align*}
\widehat{\mathcal{L}}(X,\mu) =  \lim\limits_{T\rightarrow \infty} \frac{1}{T} \sum_{t=0}^{T-1}c_\mu(x_t, u_t) ~~ \text{and} ~~
\widehat{J}_c(X)  = \lim\limits_{T\rightarrow \infty} \frac{1}{T}  \sum_{t=0}^{T-1}(4x_{t}^{\top} QWQ x_{t} + 4x_{t}^{\top}QM_3),
\end{align*}
respectively. In practice, $T$ is selected to be sufficiently large since the estimation error in the cost decreases quickly as $T \rightarrow \infty$~\citep{malik2019derivative}. Clearly, the oracle is weaker than the commonly assumed state-input trajectories in the model-free setting.

\begin{algorithm2e}[t]
	\caption{Random search algorithm to solve (\ref{prim_iterate})}
	\label{alg:learning}	
	\KwIn{Initial policy $X_0$, number of iterations $N$, smoothing radius $r$, step size $\eta$, multiplier $\mu$.}
	\For{$i=0,1,\cdots , N-1$}
	{Sample a perturbation $U\in \mathbb{R}^{m\times n}$ uniformly form a unit ball and apply $X = X_{(i)} + rU$\;
		Obtain a noisy Lagrangian function $\widehat{\mathcal{L}}(X,\mu)$ from the oracle\;
		Compute the stochastic gradient $\widehat{\nabla \mathcal{L}} = \widehat{\mathcal{L}}(X,\mu)\frac{n}{r^2}U$\;
		Update $X_{(i+1)} = X_{(i)} - \eta \widehat{\nabla \mathcal{L}}$\;}
\end{algorithm2e}

We develop a stochastic zero-order algorithm to solve (\ref{prim_iterate}) in Algorithm \ref{alg:learning}. The difficulties in the convergence analysis of Algorithm \ref{alg:learning} hinge on that (a) both the objective function $\mathcal{L}(X,\mu)$ and its feasible set $\mathcal{S}$ are non-convex; (b) unlike in~\cite{fazel2018global}, the gradient dominance property does not hold globally; (c) $\mathcal{L}(X,\mu)$ is infinite for $X \notin \mathcal{S}$ therefore the step size $\eta$ must be chosen carefully. Motivated by~\citet{malik2019derivative}, we address these problems by analysing over a compact set and showing that the algorithm remains in it with large probabilities.

Define the gap between an initial policy $X_{(0)}$ and $X^*(\mu)$ as $\Delta_0 = \mathcal{L}(X_{(0)},\mu) - \mathcal{L}(X^*(\mu),\mu)$ as well as the compact set
\begin{equation}\label{def:initial set}
\mathcal{S}_0=\left\{X \mid \mathcal{L}(X,\mu)-\mathcal{L}(X_{(0)},\mu) \leq 10 \Delta_{0}\right\} \subset \mathcal{S}.
\end{equation}
We denote the gradient dominance constant over $\mathcal{S}_0$ as $\lambda_0$. Also, define the Lipschitz constants
\begin{equation}
\beta_0 = \sup_{X \in \mathcal{S}_0} \beta_X,~~~~~~~~
\zeta_0 = \sup_{X \in \mathcal{S}_0} \zeta_X,~~~~~~~~
\gamma_0 = \inf_{X \in \mathcal{S}_0} \gamma_X.
\end{equation}
By doing so, the properties hold globally on $\mathcal{S}_0$. To ensure the step size $\eta$ not too large, it should be set according to the variance of the estimated gradient $\widehat{\nabla \mathcal{L}}$. The gradient norm is defined as
\begin{equation}
G_{\infty}=\sup _{X \in \mathcal{S}_0}\|\widehat{\nabla \mathcal{L}}\|_{2}, \quad \text { and } ~~ G_{2}=\sup _{X \in \mathcal{S}_0} \mathbb{E}\|\widehat{\nabla \mathcal{L}}-\mathbb{E}[\widehat{\nabla \mathcal{L}} | X]\|_{2}^{2}.
\end{equation}
We make the following assumption for the noise $w_t$ to ensure the existence of the gradient norm.
\begin{assum}\label{asum}
	The noise $w_t$ is uniformly bounded, i.e., $\|w_t\| \leq v$ where $v>0$ is a constant.
\end{assum}

The following theorem shows that with a large probability, Algorithm \ref{alg:learning} converges and $\{X_{(i)}\}$ always remain in $\mathcal{S}_0$. For notional convenience, we denote $\theta_0 = \min \{\frac{1}{2\beta_0},\frac{\gamma_0}{\zeta_0}\}$. 
\begin{theorem}\label{theorem:learning}
	Suppose that the step-size and smoothing radius are chosen such that
	\begin{align*}
	\eta \leq \min \{ \frac{\epsilon}{240 \lambda_0 \beta_0 G_2}, \frac{1}{2\beta_0}, \frac{\gamma_0}{G_{\infty}}   \} ~~~~\text{and} ~~~~
	r  \leq \min \{ \frac{\theta_0}{8 \lambda_0 \beta_0 }\sqrt{\frac{\epsilon}{15}}, \frac{1}{2\beta_0} \sqrt{\frac{\epsilon}{30 \lambda_0}}, \gamma_0 \}.
	\end{align*}
	Then, for any error tolerance $\epsilon$ such that $\epsilon \log \left(120 \Delta_{0} / \epsilon\right)<\frac{10}{3} \Delta_{0}$ and $N=\frac{4 \lambda_0}{\eta} \log \left(\frac{120 \Delta_{0}}{\epsilon}\right)$, with probability greater than $\frac{3}{4}$ the iterations in Algorithm \ref{alg:learning} yield a controller $X_N$ such that 
	$$
	\mathcal{L}(X_{(N)}, \mu)- \mathcal{L}(X^*, \mu) \leq \epsilon.
	$$
\end{theorem}
The proof follows from Theorem 1 in ~\citet{malik2019derivative} and we extend it in that we derive the gradient norm bound for the average cost setting. In view of \citet{furieri2020learning}, the convergence probability in Theorem \ref{theorem:learning} can be improved to $1-\delta$ for any $0<\delta<1$ by working on
$
\mathcal{S}_{\delta}=\{X \mid \mathcal{L}(X,\mu)-\mathcal{L}(X_{(0)},\mu) \leq 10 \delta^{-1} \Delta_{0}\} \subset \mathcal{S}.
$
For simplicity, we adopt the methodology in \cite{malik2019derivative}.

\begin{algorithm2e}[t]
	\caption{Primal-dual learning algorithm for the risk-constrained LQR}
	\label{alg:dual}
	\KwIn{Initial multiplier $\mu_1$, step size ${\xi_j}$, $j \in \{1,2,\dots\}$.}
	\For{$j=1,2,\dots$}
	{\textbf{Step 1: learning the dual function}\\
		Learn a policy $X_j\in \mathop{\text{argmin}} \limits_{X \in \mathcal{S}}~ \mathcal{L}(X,\mu_j)$ by Algorithm \ref{alg:learning}\;
		\textbf{Step 2: dual ascent}\\
		Obtain a noisy sample $\widehat{J_c}(X_j)$ from the oracle\;
		Estimate the subgradient $\hat{\omega}(\mu_j)$ by (\ref{def:subgradient})\;
		Update the dual variable by $\mu_{j+1} = [\mu_{j} + \xi_j \hat{\omega}(\mu_j)]_{+}$\;}	
\end{algorithm2e}

\subsection{Primal-dual Algorithm}
By dual theory~\citep{nesterov2013introductory,nedic2009subgradient}, the subgradient of $D(\mu)$ is given as 
\begin{equation}\label{def:sub}
\omega(\mu) = J_c(X^*(\mu)) - \bar{\rho},
\end{equation}
However, $J_c(X^*(\mu))$ cannot be computed directly as we do not have a dynamical model. To this end, we estimate it by a noisy sample from the oracle. The subgradient $\omega(\mu)$ is approximated as
\begin{equation}\label{def:subgradient}
\widehat{\omega}(\mu)=  \widehat{J_c}(X^*(\mu)) - \bar{\rho}
\end{equation}

We present our complete primal-dual algorithm in Algorithm \ref{alg:dual}. In general, there is no guarantee that a primal variable sequence will converge to the optimal solution unless the subdifferential at the dual variables is a singleton~\citep{bertsekas1997nonlinear,boyd2004convex}. Fortunately, this is indeed the case for (\ref{prob:new_rclqr}) as minimizing the Lagrangian function yields a unique solution, which implies that the subgradient in (\ref{def:sub}) is actually a gradient. Furthermore, we analyze its convergence by leveraging the boundedness of the gradient norm $\|\hat{\omega}(\mu)\|$, which is evidenced by the fact that a stabilizing policy $X^*(\mu)$ yields a finite cost. 
\begin{theorem}\label{theorem:primal-dual}
	Let $\mathbb{E}\|\hat{\omega}(\mu_j)\| \leq b$ and $\mathbb{E}\|\mu_j\| \leq e$ with $b,e>0$. Define $\bar{\mu}_j = \frac{1}{j} \sum_{i=1}^j \mu_{i}$. Then, by selecting a diminishing step size $\xi_j = \frac{1}{be}\sqrt \frac{2}{j}$, Algorithm \ref{alg:dual} satisfies
	$$
	D^*-\mathbb{E}[D(\bar{\mu}_j)] \leq \frac{3be}{\sqrt{j}}.
	$$
\end{theorem}

\section{Simulation Results}

%\begin{figure*}[t]
%	\centering
%	\subfigure[ ]{
%		\label{pic:lag_cost}
%		\includegraphics[width=0.3 \textwidth]{lag_cost.pdf}
%	}
%	\subfigure[ ]{
%		\label{pic:risk_cost}
%		\includegraphics[width=0.3 \textwidth]{pic_dual_cost.pdf}
%	}
%	\subfigure[ ]{
%		\label{pic:multiplier}
%		\includegraphics[width=0.3 \textwidth]{multiplier.pdf}
%	}
%	\caption{{ (a) The Lagrangian function converges from $377$ to its optimal value $107$; (b) $J_c(X^*(\mu))$ decreases until the risk constraint is satisfied; (c) the multiplier converges to the optimal value.
%	}}
%\end{figure*}
\begin{figure}[t]
	\centerline{\includegraphics[width=70mm]{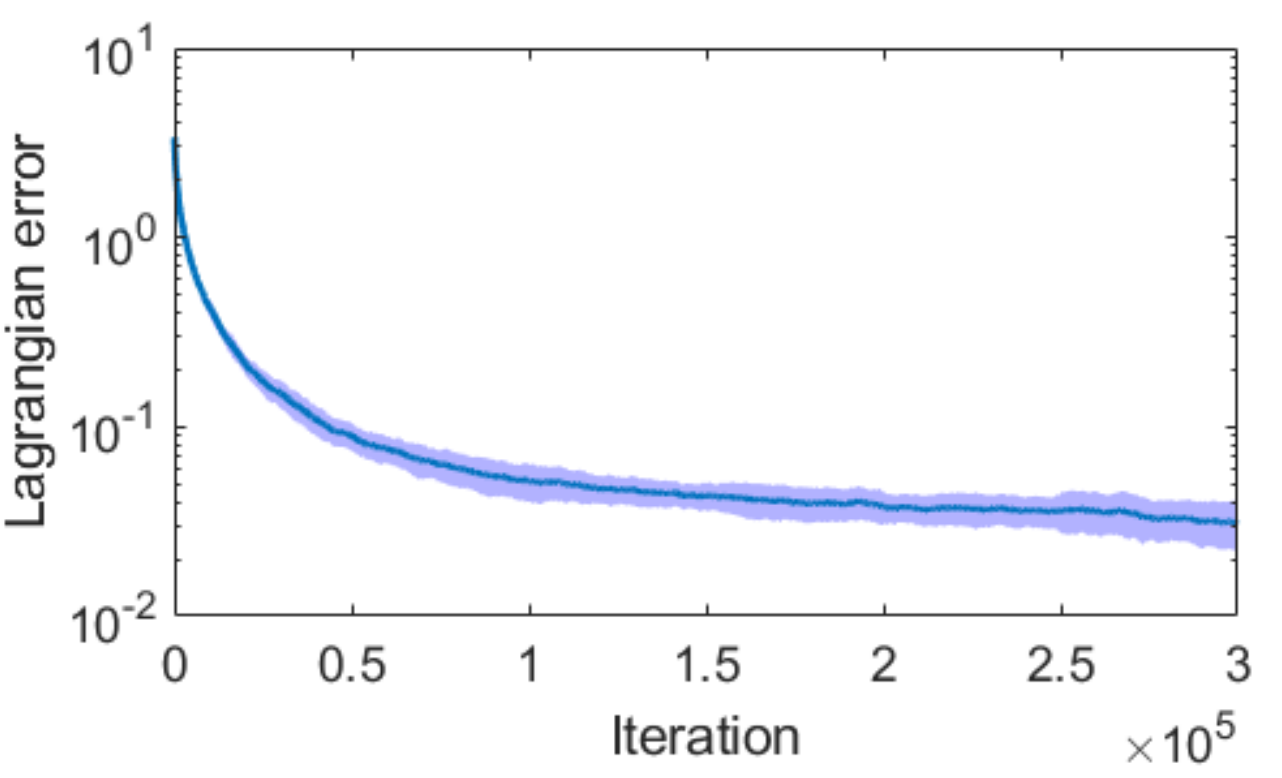}}
	\caption{Relative Lagrangian error $(\mathcal{L}(X_i,\mu) - D(\mu))/D(\mu)$ of Algorithm \ref{alg:learning} for a fixed $\mu$. The bold centreline denotes the mean of 20 trials and the shaded region demonstrates their standard deviation.}
	\label{pic:random_search}
\end{figure}

\begin{figure}[t]
	\centering
	\subfigure[Optimality gap $|J(X_j)- J(X^*)|/J(X^*)$.]{
		\includegraphics[width=70mm]{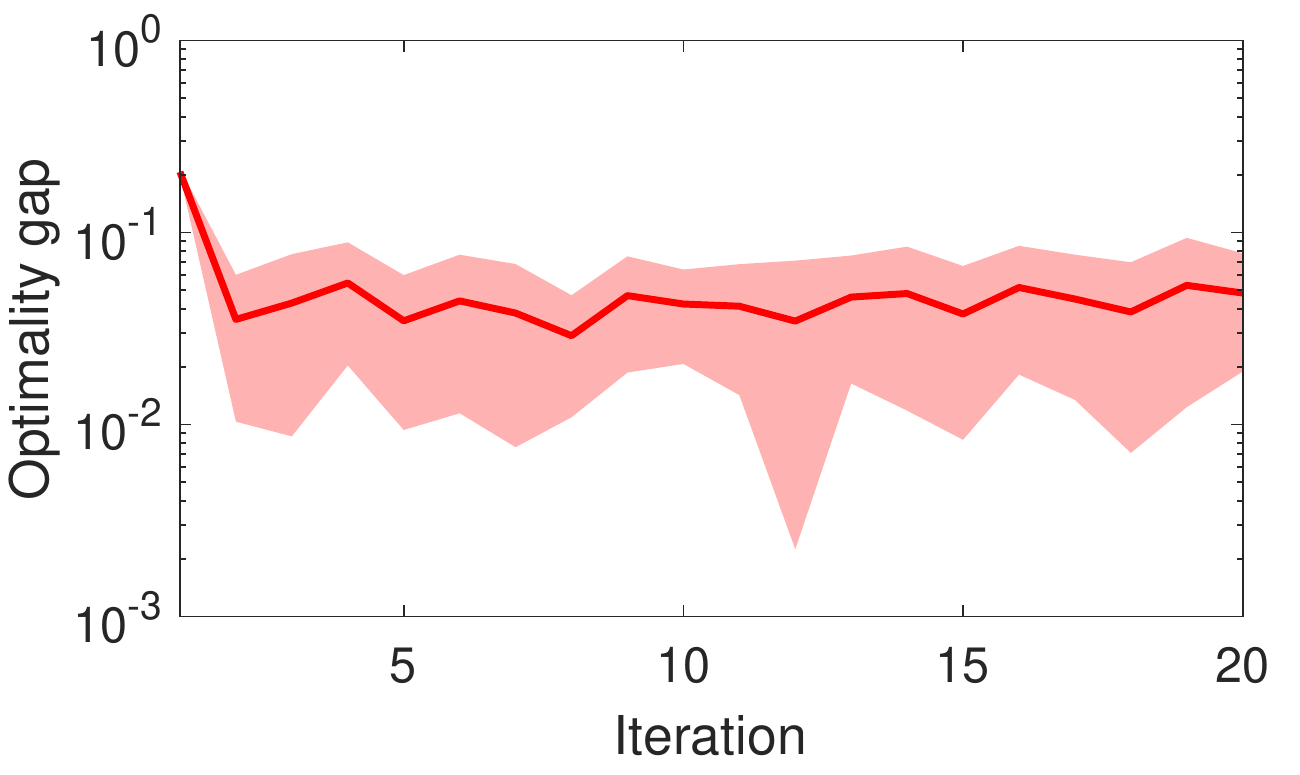}}
	\subfigure[Risk constraint violation $|J_c(X_j)- \bar{\rho}|/\bar{\rho}$.]{
		\includegraphics[width=70mm]{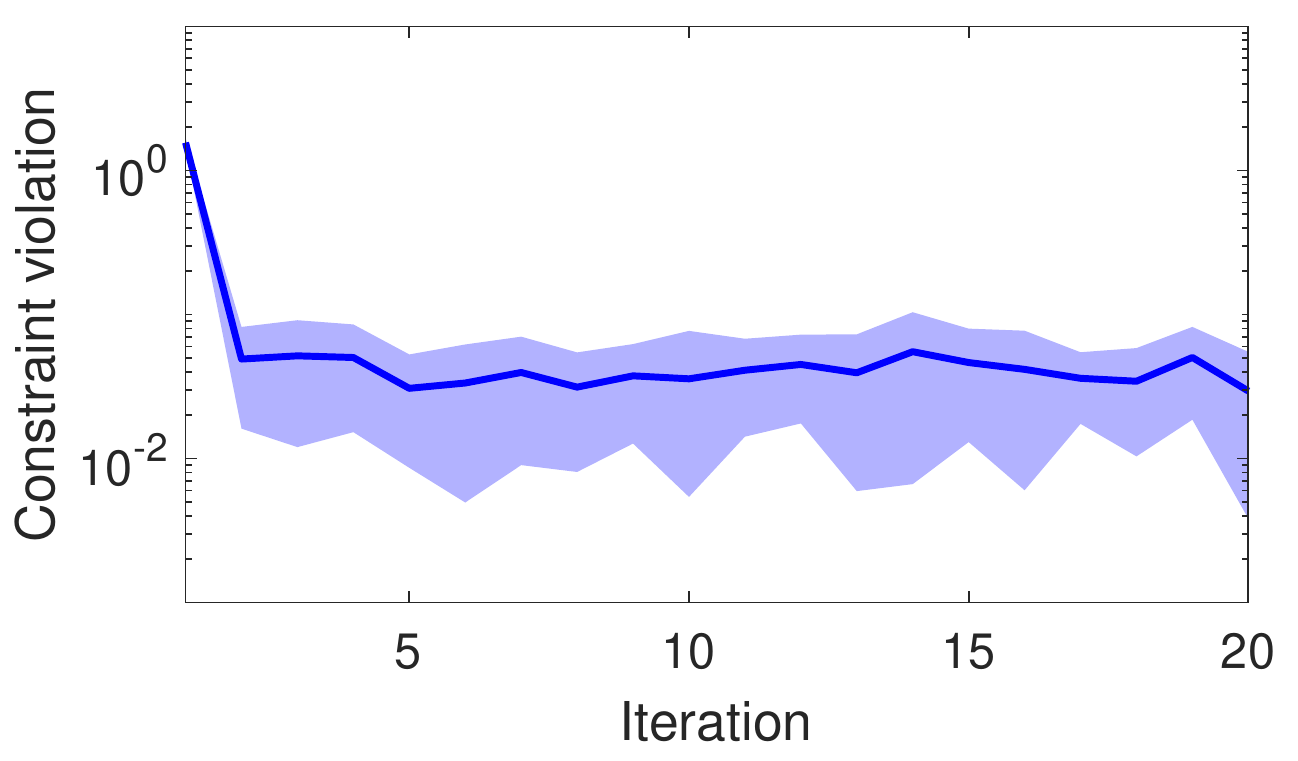}}
	\caption{Convergence of the primal-dual learning method. The centreline denotes the mean of 20 trials and the shaded region demonstrates their standard deviation.}
	\label{pic:pd_mf}
\end{figure}
In the experiment, we consider an unmanned aerial vehicle (UAV) that operates in a 2-D x-y plane. The discrete-time dynamical model is given by a double integrator as
\begin{equation}\label{def:model}
x_{k+1}=\begin{bmatrix}
1 & 0.5 & 0 & 0 \\
0 & 1 & 0 & 0 \\
0 & 0 & 1 & 0.5 \\
0 & 0 & 0 & 1
\end{bmatrix} x_{k}+\begin{bmatrix}
0.125 & 0 \\
0.5 & 0 \\
0 & 0.125 \\
0 & 0.5
\end{bmatrix}\left(u_{k}+w_{k}\right),
\end{equation}
where $(x_{k,1}, x_{k,3})$ and $(x_{k,2}, x_{k,4})$ denote the position and velocity, respectively, $u_k$ represents the acceleration and $w_k$ is the input disturbance from the wind. Suppose that the gust $w_{k,1}$ in the direction of $x_{k,1}$ is subject to a mixed Gaussian distribution of $\mathcal{N}(3,30)$ and $\mathcal{N}(8,60)$ with weights 0.2 and 0.8, respectively. In contrast, the gust $w_{k,2}$ in the orthogonal direction satisfies $w_{k,2} \sim \mathcal{N}(0,0.01)$. We set the penalty matrix in (\ref{prob:new_rclqr}) as
$
Q = \text{diag}(1,0.1,2,0.2)$ and $R = \text{diag}(1,1)
$. 

We verify the convergence of the proposed primal-dual learning method by examining the optimality gap and the risk constraint violation. Since the system (\ref{def:model}) is open-loop unstable, we select an initial policy
$$
K_0 = \begin{bmatrix}
0.5 & 0.5 & 0 & 0 \\
0 & 0 & 0.5 & 0.5
\end{bmatrix} ~~\text{and}~~
l_0 = \begin{bmatrix}
-6\\
0
\end{bmatrix},
$$
which is readily stabilizing. The risk tolerance in (\ref{prob:new_rclqr}) is set as $\bar{\rho} = 15$.

We first perform the random search in Algorithm \ref{alg:learning} to solve (\ref{prim_iterate}). We set the initial multiplier as $\mu_1 = 0$, the smoothing radius as $r = 0.2$ and the sample horizon of the oracle as $T=100$. We empirically select the step size $\eta = 1\times10^{-5}$ to yield a good performance, which is common in the policy optimization of LQR problems. We perform 20 independent trials and display the relative Lagrangian error when $\mu = 2$ in Fig. \ref{pic:random_search}. Clearly, Algorithm \ref{alg:learning} converges to relative error $3\%$ within $3\times10^{5}$ iterations and exhibits small variance.

Then, we conduct our primal-dual learning method in Algorithm \ref{alg:dual} with 20 independent trials. The horizon of the risk oracle is set as $T = 10^4$ to reduce the variance of subgradient. Denote the optimal value of (\ref{prob:new_rclqr}) as $J(X^*)$. Since there is an inevitable error in the Lagrangian (around $3\%$) per iteration, the optimality gap and constraint violation finally converge to $5\%$, see Fig. \ref{pic:pd_mf}. The variance of them originates from the primal iteration and the subgradient estimation.

\section{Conclusion}

In this paper, we have proposed a primal-dual learning framework for the model-free risk-constrained LQR. In particular, we have shown that the Lagrangian function is both locally gradient dominated and Lipschitz, based on which the strong duality is established. Furthermore, we have shown the global convergence of the proposed primal-dual learning algorithm. 

This work only considers the gradient descent method in a stochastic form. However, the optimization landscape of natural gradient and Gauss-Newton method for the risk-constrained LQR, even in the model-based setting, is still unclear. We have considered the policy gradient primal-dual method in the model-based setting in \cite{zhao2021global}. Also note that there is only one constraint in our optimization problem. It is also interesting to study the PO for LQR with multiple constraints, which will be our future work.

\acks{We would like to thank Mr. Kaiqing Zhang from University of Illinois at Urbana-Champaign for his constructive suggestions, and Mr. Jiaqi Zhang from Tsinghua University for his advice on writing. This research was supported by National Natural Science Foundation of China under Grant no. 62033006.}	
\bibliography{mybibfile}

\begin{thebibliography}{41}
\providecommand{\natexlab}[1]{#1}
\providecommand{\url}[1]{\texttt{#1}}
\expandafter\ifx\csname urlstyle\endcsname\relax
  \providecommand{\doi}[1]{doi: #1}\else
  \providecommand{\doi}{doi: \begingroup \urlstyle{rm}\Url}\fi

\bibitem[{\AA}str{\"o}m(2012)]{aastrom2012introduction}
Karl~J {\AA}str{\"o}m.
\newblock \emph{Introduction to stochastic control theory}.
\newblock Courier Corporation, 2012.

\bibitem[B{\"a}uerle and Rieder(2014)]{bauerle2014more}
Nicole B{\"a}uerle and Ulrich Rieder.
\newblock More risk-sensitive markov decision processes.
\newblock \emph{Mathematics of Operations Research}, 39\penalty0 (1):\penalty0
  105--120, 2014.

\bibitem[Bertsekas(1995)]{bertsekas1995dynamic}
Dimitri~P Bertsekas.
\newblock \emph{Dynamic programming and optimal control}, volume~1.
\newblock Athena scientific Belmont, MA, 1995.

\bibitem[Bertsekas(1997)]{bertsekas1997nonlinear}
Dimitri~P Bertsekas.
\newblock Nonlinear programming.
\newblock \emph{Journal of the Operational Research Society}, 48\penalty0
  (3):\penalty0 334--334, 1997.

\bibitem[Bertsekas(2019)]{bertsekas2019reinforcement}
Dimitri~P Bertsekas.
\newblock \emph{Reinforcement learning and optimal control}.
\newblock Athena Scientific, 2019.

\bibitem[Borkar and Jain(2014)]{borkar2014risk}
Vivek Borkar and Rahul Jain.
\newblock Risk-constrained markov decision processes.
\newblock \emph{IEEE Transactions on Automatic Control}, 59\penalty0
  (9):\penalty0 2574--2579, 2014.

\bibitem[Boyd et~al.(2004)Boyd, Boyd, and Vandenberghe]{boyd2004convex}
Stephen Boyd, Stephen~P Boyd, and Lieven Vandenberghe.
\newblock \emph{Convex optimization}.
\newblock Cambridge university press, 2004.

\bibitem[Bu et~al.(2019)Bu, Mesbahi, Fazel, and Mesbahi]{bu2019lqr}
Jingjing Bu, Afshin Mesbahi, Maryam Fazel, and Mehran Mesbahi.
\newblock {LQR} through the lens of first order methods: Discrete-time case.
\newblock \emph{arXiv preprint arXiv:1907.08921}, 2019.

\bibitem[Chapman et~al.(2019)Chapman, Lacotte, Tamar, Lee, Smith, Cheng, Fisac,
  Jha, Pavone, and Tomlin]{chapman2019risk}
Margaret~P Chapman, Jonathan Lacotte, Aviv Tamar, Donggun Lee, Kevin~M Smith,
  Victoria Cheng, Jaime~F Fisac, Susmit Jha, Marco Pavone, and Claire~J Tomlin.
\newblock A risk-sensitive finite-time reachability approach for safety of
  stochastic dynamic systems.
\newblock In \emph{American Control Conference}, pages 2958--2963, 2019.

\bibitem[Chow et~al.(2017)Chow, Ghavamzadeh, Janson, and Pavone]{chow2017risk}
Yinlam Chow, Mohammad Ghavamzadeh, Lucas Janson, and Marco Pavone.
\newblock Risk-constrained reinforcement learning with percentile risk
  criteria.
\newblock \emph{The Journal of Machine Learning Research}, 18\penalty0
  (1):\penalty0 6070--6120, 2017.

\bibitem[Fazel et~al.(2018)Fazel, Ge, Kakade, and Mesbahi]{fazel2018global}
Maryam Fazel, Rong Ge, Sham Kakade, and Mehran Mesbahi.
\newblock Global convergence of policy gradient methods for the linear
  quadratic regulator.
\newblock In \emph{International Conference on Machine Learning}, pages
  1467--1476, 2018.

\bibitem[Furieri et~al.(2020)Furieri, Zheng, and
  Kamgarpour]{furieri2020learning}
Luca Furieri, Yang Zheng, and Maryam Kamgarpour.
\newblock Learning the globally optimal distributed {LQ} regulator.
\newblock In \emph{Learning for Dynamics and Control}, pages 287--297, 2020.

\bibitem[{Gravell} et~al.(2020){Gravell}, {Mohajerin Esfahani}, and
  {Summers}]{gravell2020learn}
B.~{Gravell}, P.~{Mohajerin Esfahani}, and T.~H. {Summers}.
\newblock Learning optimal controllers for linear systems with multiplicative
  noise via policy gradient.
\newblock \emph{IEEE Transactions on Automatic Control}, 2020.
\newblock ISSN 1558-2523.
\newblock \doi{10.1109/TAC.2020.3037046}.

\bibitem[Horn and Johnson(2012)]{horn2012matrix}
Roger~A Horn and Charles~R Johnson.
\newblock \emph{Matrix analysis}.
\newblock Cambridge university press, 2012.

\bibitem[Jacobson(1973)]{jacobson1973optimal}
David Jacobson.
\newblock Optimal stochastic linear systems with exponential performance
  criteria and their relation to deterministic differential games.
\newblock \emph{IEEE Transactions on Automatic control}, 18\penalty0
  (2):\penalty0 124--131, 1973.

\bibitem[Jansch-Porto et~al.(2020)Jansch-Porto, Hu, and
  Dullerud]{jansch2020convergence}
Joao~Paulo Jansch-Porto, Bin Hu, and Geir Dullerud.
\newblock Convergence guarantees of policy optimization methods for markovian
  jump linear systems.
\newblock \emph{arXiv preprint arXiv:2002.04090}, 2020.

\bibitem[Li et~al.(2019)Li, Tang, Zhang, and Li]{li2019distributed}
Yingying Li, Yujie Tang, Runyu Zhang, and Na~Li.
\newblock Distributed reinforcement learning for decentralized linear quadratic
  control: A derivative-free policy optimization approach.
\newblock \emph{arXiv preprint arXiv:1912.09135}, 2019.

\bibitem[Lillicrap et~al.(2016)Lillicrap, Hunt, Pritzel, Heess, Erez, Tassa,
  Silver, and Wierstra]{lillicrap2016continuous}
Timothy~P Lillicrap, Jonathan~J Hunt, Alexander Pritzel, Nicolas Heess, Tom
  Erez, Yuval Tassa, David Silver, and Daan Wierstra.
\newblock Continuous control with deep reinforcement learning.
\newblock In \emph{International Conference on Learning Representations}, 2016.

\bibitem[Malik et~al.(2019)Malik, Pananjady, Bhatia, Khamaru, Bartlett, and
  Wainwright]{malik2019derivative}
Dhruv Malik, Ashwin Pananjady, Kush Bhatia, Koulik Khamaru, Peter Bartlett, and
  Martin Wainwright.
\newblock Derivative-free methods for policy optimization: Guarantees for
  linear quadratic systems.
\newblock In \emph{The 22nd International Conference on Artificial Intelligence
  and Statistics}, pages 2916--2925, 2019.

\bibitem[Mnih et~al.(2015)Mnih, Kavukcuoglu, Silver, Rusu, Veness, Bellemare,
  Graves, Riedmiller, Fidjeland, Ostrovski, et~al.]{mnih2015human-level}
Volodymyr Mnih, Koray Kavukcuoglu, David Silver, Andrei~A Rusu, Joel Veness,
  Marc~G Bellemare, Alex Graves, Martin Riedmiller, Andreas~K Fidjeland, Georg
  Ostrovski, et~al.
\newblock Human-level control through deep reinforcement learning.
\newblock \emph{Nature}, 518\penalty0 (7540):\penalty0 529--533, 2015.

\bibitem[Mohammadi et~al.(2020)Mohammadi, Soltanolkotabi, and
  Jovanovi{\'c}]{mohammadi2020linear}
Hesameddin Mohammadi, Mahdi Soltanolkotabi, and Mihailo~R Jovanovi{\'c}.
\newblock On the linear convergence of random search for discrete-time {LQR}.
\newblock \emph{IEEE Control Systems Letters}, 5\penalty0 (3):\penalty0
  989--994, 2020.

\bibitem[Moore et~al.(1997)Moore, Elliott, and Dey]{moore1997risk}
John~B Moore, Robert~J Elliott, and Subhrakanti Dey.
\newblock Risk-sensitive generalizations of minimum variance estimation and
  control.
\newblock \emph{Journal of Mathematical Systems Estimation and Control},
  7:\penalty0 123--126, 1997.

\bibitem[Nedi{\'c} and Ozdaglar(2009)]{nedic2009subgradient}
Angelia Nedi{\'c} and Asuman Ozdaglar.
\newblock Subgradient methods for saddle-point problems.
\newblock \emph{Journal of optimization theory and applications}, 142\penalty0
  (1):\penalty0 205--228, 2009.

\bibitem[Nesterov(2013)]{nesterov2013introductory}
Yurii Nesterov.
\newblock \emph{Introductory lectures on convex optimization: A basic course},
  volume~87.
\newblock Springer Science \& Business Media, 2013.

\bibitem[Pan et~al.(2019)Pan, Seita, Gao, and Canny]{pan2019risk}
Xinlei Pan, Daniel Seita, Yang Gao, and John Canny.
\newblock Risk averse robust adversarial reinforcement learning.
\newblock In \emph{International Conference on Robotics and Automation}, pages
  8522--8528, 2019.

\bibitem[Paternain et~al.(2019)Paternain, Chamon, Calvo-Fullana, and
  Ribeiro]{paternain2019constrained}
Santiago Paternain, Luiz Chamon, Miguel Calvo-Fullana, and Alejandro Ribeiro.
\newblock Constrained reinforcement learning has zero duality gap.
\newblock In \emph{Advances in Neural Information Processing Systems}, pages
  7555--7565, 2019.

\bibitem[Prashanth~L and Fu(2018)]{prashanth2018risk}
A~Prashanth~L and Michael Fu.
\newblock Risk-sensitive reinforcement learning: A constrained optimization
  viewpoint.
\newblock \emph{arXiv}, pages arXiv--1810, 2018.

\bibitem[Rockafellar et~al.(2000)Rockafellar, Uryasev,
  et~al.]{rockafellar2000optimization}
R~Tyrrell Rockafellar, Stanislav Uryasev, et~al.
\newblock Optimization of conditional value-at-risk.
\newblock \emph{Journal of risk}, 2:\penalty0 21--42, 2000.

\bibitem[Roulet et~al.(2020)Roulet, Fazel, Srinivasa, and
  Harchaoui]{roulet2020convergence}
Vincent Roulet, Maryam Fazel, Siddhartha Srinivasa, and Zaid Harchaoui.
\newblock On the convergence of the iterative linear exponential quadratic
  gaussian algorithm to stationary points.
\newblock In \emph{American Control Conference}, pages 132--137, 2020.

\bibitem[Sopasakis et~al.(2019)Sopasakis, Schuurmans, and
  Patrinos]{sopasakis2019risk}
Pantelis Sopasakis, Mathijs Schuurmans, and Panagiotis Patrinos.
\newblock Risk-averse risk-constrained optimal control.
\newblock In \emph{2019 18th European Control Conference (ECC)}, pages
  375--380. IEEE, 2019.

\bibitem[Speyer et~al.(1992)Speyer, Fan, and Banavar]{speyer1992optimal}
Jason~L Speyer, C-H Fan, and Ravi~N Banavar.
\newblock Optimal stochastic estimation with exponential cost criteria.
\newblock In \emph{Proceedings of the 31st IEEE Conference on Decision and
  Control}, pages 2293--2299, 1992.

\bibitem[Sutton et~al.(1998)Sutton, Barto, et~al.]{sutton1998introduction}
Richard~S Sutton, Andrew~G Barto, et~al.
\newblock \emph{Introduction to reinforcement learning}, volume 135.
\newblock MIT press Cambridge, 1998.

\bibitem[Tessler et~al.(2018)Tessler, Mankowitz, and Mannor]{tessler2018reward}
Chen Tessler, Daniel~J Mankowitz, and Shie Mannor.
\newblock Reward constrained policy optimization.
\newblock \emph{arXiv preprint arXiv:1805.11074}, 2018.

\bibitem[{Tsiamis} et~al.(2020){Tsiamis}, {Kalogerias}, {Chamon}, {Ribeiro},
  and {Pappas}]{tsiamis2020risk}
A.~{Tsiamis}, D.~S. {Kalogerias}, L.~F.~O. {Chamon}, A.~{Ribeiro}, and G.~J.
  {Pappas}.
\newblock Risk-constrained linear-quadratic regulators.
\newblock In \emph{59th IEEE Conference on Decision and Control (CDC)}, pages
  3040--3047, 2020.
\newblock \doi{10.1109/CDC42340.2020.9303967}.

\bibitem[Wen and Topcu(2018)]{wen2018constrained}
Min Wen and Ufuk Topcu.
\newblock Constrained cross-entropy method for safe reinforcement learning.
\newblock In \emph{Advances in Neural Information Processing Systems}, pages
  7450--7460, 2018.

\bibitem[Whittle(1981)]{whittle1981risk}
Peter Whittle.
\newblock Risk-sensitive linear quadratic gaussian control.
\newblock \emph{Advances in Applied Probability}, pages 764--777, 1981.

\bibitem[Yu et~al.(2019)Yu, Yang, Kolar, and Wang]{yu2019convergent}
Ming Yu, Zhuoran Yang, Mladen Kolar, and Zhaoran Wang.
\newblock Convergent policy optimization for safe reinforcement learning.
\newblock In \emph{Advances in Neural Information Processing Systems}, pages
  3127--3139, 2019.

\bibitem[Zhang et~al.(2019)Zhang, Yang, and Basar]{zhang2019policy}
Kaiqing Zhang, Zhuoran Yang, and Tamer Basar.
\newblock Policy optimization provably converges to nash equilibria in zero-sum
  linear quadratic games.
\newblock In \emph{Advances in Neural Information Processing Systems}, pages
  11598--11610, 2019.

\bibitem[Zhang et~al.(2020)Zhang, Hu, and Basar]{zhang2020policy}
Kaiqing Zhang, Bin Hu, and Tamer Basar.
\newblock Policy optimization for $\mathcal{H}_2$ linear control with
  $\mathcal{H}_{\infty}$ robustness guarantee: Implicit regularization and
  global convergence.
\newblock In \emph{Learning for Dynamics and Control}, pages 179--190, 2020.

\bibitem[Zhao et~al.(2021{\natexlab{a}})Zhao, You, and
  Ba\c{s}ar]{zhao2021infinitehorizon}
Feiran Zhao, Keyou You, and Tamer Ba\c{s}ar.
\newblock Infinite-horizon risk-constrained linear quadratic regulator with
  average cost.
\newblock \emph{arXiv preprint arXiv:2103.15363}, 2021{\natexlab{a}}.

\bibitem[Zhao et~al.(2021{\natexlab{b}})Zhao, You, and
  Ba{\c{s}}ar]{zhao2021global}
Feiran Zhao, Keyou You, and Tamer Ba{\c{s}}ar.
\newblock Global convergence of policy gradient primal-dual methods for
  risk-constrained {LQRs}.
\newblock \emph{arXiv preprint arXiv:2104.04901}, 2021{\natexlab{b}}.

\end{thebibliography}

\begin{appendices}
\section{Proof in Section \ref{subsec:closed-form}}
In this section, we establish the closed-form expression for the Lagrangian function $\mathcal{L}(X,\mu)$ and its gradient.

We define the value function with the reshaped cost $c_\mu(x_t,u_t)$ as
\begin{equation}\label{def:value}
V_{K,l}(x)=\mathbb{E} \sum_{t = 0}^{\infty} [c_\mu(x_{t}, u_{t}) -\mathcal{L}(K,l,\mu)| x_{0}=x],
\end{equation}
which differs from its classical definition in that it accumulates the relative cost with respect to the average cost $\mathcal{L}(K,l,\mu)$. This definition is to ensure that for a stabilizing policy the value function is finite. Suppose that $P_K \geq 0$ is the solution of the algebraic Riccati equation
$$
P_{K} = Q_{\mu} +K^{\top} R K + (A-B K)^{\top} P_{K}(A-B K)
$$
and let $E_{K}=(R+B^{\top} P_{K} B) K-B^{\top} P_{K} A$ and $V = (I - (A-BK))^{-1}$. In the following proposition, we show that $V_{K,l}(x)$ is quadratic in $x$, by the derivation of which the closed-form of $\mathcal{L}(K,l,\mu)$ can be obtained.
\begin{proposition}[Closed-form expression]\label{prop:lag}
	The value function in (\ref{def:value}) for a stabilizing policy $X \in \mathcal{S}$ is given by
	\begin{equation}\label{equ:value}
	V_{K,l}(x)= x^{\top}P_Kx + g_{K,l}^{\top}x + z_{K,l},
	\end{equation}
	where $g_{K,l}^{\top} = 2(-l^{\top}E_K+S^{\top} + \bar{w}^{\top}P_K(A-BK))V$
	and $z_{K,l}$ is a constant. Moreover, the Lagrangian function $\mathcal{L}(K,l,\mu)$ can be expressed with $P_K$ and $g_{K,l}$ as
	\begin{equation}
	\mathcal{L}(K,l,\mu) = \mathrm{tr}\{P_K(W + (Bl+\bar{w})(Bl+\bar{w})^{\top})\} + g_{K,l}^{\top}(Bl+\bar{w})  + l^{\top}Rl - \mu \bar{\rho}.
	\end{equation}
\end{proposition}

\subsection{Proof of Proposition \ref{prop:lag}}
	Inserting $u_t = -K x_t + l$ into $V_{K,l}(x)$ in (\ref{def:value}), it follows that
	$$
	V_{K,l}(x) = \sum_{t = 0}^{\infty}\mathbb{E}[ x_t^{\top}(Q_{\mu}+K^{\top}RK)x_t + (2S^{\top}-2l^{\top}RK)x_t + l^{\top}Rl - \mu \bar{\rho}
	]  -\mathcal{L}(K, l, \mu).
	$$
	Due to the linear dynamical model (\ref{def:model}), it follows immediately from the dynamic programming theory \citep{bertsekas1995dynamic} that $V_{K,l}(x)$ has a quadratic form, i.e.,
	$$
	V_{K,l}(x)= x^{\top}P_Kx + g_{K,l}^{\top}x + z_{K,l},
	$$
	where $P_K,  g_{K,l}, z_{K,l}$ are parameters to be determined. Clearly, $V_{K,l}(x)$ satisfies the Bellman equation
	$$
	V_{K,l}(x_t) = x_t^{\top}(Q_{\mu}+K^{\top}RK)x_t + (2S^{\top}-2l^{\top}RK)x_t + l^{\top}Rl - \mathcal{L}(K, l, \mu) + \mathbb{E}[V_{K,l}(x_{t+1})].
	$$
	Thus, it follows that
	\begin{align*}
	&~~~~~x_t^{\top}P_Kx_t + g_{K,l}^{\top}x_t + z_{K,l}\\
	& = x_t^{\top}(Q_{\mu}+K^{\top}RK)x_t + (2S^{\top}-2l^{\top}RK)x_t + l^{\top}Rl - \mathcal{L}(K, l, \mu)- \mu \bar{\rho} \\
	&~~~~+ \mathbb{E}  [(A-BK)x_t +Bl + w_t]^{\top}P_K [(A-BK)x_t +Bl + w_t] + \mathbb{E}[g_{K,l}^{\top}x_t] + z_{K,l}
	\\
	&= x_t^{\top}[Q_{\mu} + K^{\top}RK+(A-B K)^{\top} P_{K}(A-B K)]x_t + 2[-l^{\top}E_K+S^{\top} + \bar{w}^{\top}P_K(A-BK)]x_t \\
	&~~~~+ \mathrm{tr}\bigl[P_K(W + (Bl+\bar{w})(Bl+\bar{w})^{\top} )\bigr] + g_{K,l}^{\top}(Bl+\bar{w})  + l^{\top}Rl-\mathcal{L}(K, l, \mu) - \mu \bar{\rho} + z_{K,l},
	\end{align*}
	which holds for all $x_t \in \mathbb{R}^n$. Hence, we can solve the parameters 
	\begin{align*}
	P_{K} &= Q_K+(A-B K)^{\top} P_{K}(A-B K), \\
	g_{K,l}^{\top} &= 2[-l^{\top}E_K+S^{\top} + \bar{w}^{\top}P_K(A-BK)]V,
	\end{align*}
	and obtain
	$$
	\mathcal{L}(K,l,\mu) = \mathrm{tr}\{P_K(W + (Bl+\bar{w})(Bl+\bar{w})^{\top})\} + g_{K,l}^{\top}(Bl+\bar{w})  + l^{\top}Rl- \mu \bar{\rho}.
	$$
\subsection{Proof of Proposition \ref{prop:grad}}
Before deriving the gradient expression $\nabla \mathcal{L}(X,\mu)$, we first present the following lemma.
\begin{lemma}
	Define $Q_K = Q_\mu + K^{\top}RK$. The Lagrangian function and its gradient can also be expressed as
	\begin{equation}\notag
	\begin{aligned}
	\mathcal{L}(K,l,\mu) &= \mathrm{tr}\{Q_K(\Sigma_{K} + \bar{x}_{K,l}\bar{x}_{K,l}^{\top})\} + (2S^{\top} -2l^{\top}RK)\bar{x}_{K,l} + l^{\top}Rl- \mu \bar{\rho} \\
	\nabla_{K} \mathcal{L}(K,l,\mu) &= 2E_K\Sigma_{K} - \nabla_{l}  \mathcal{L}(K,l,\mu) \bar{x}_{K,l}^{\top}\\
	\nabla_{l} \mathcal{L}(K,l,\mu) & = 2B^{\top}V^{\top}(Q_K\bar{x}_{K,l} -K^{\top}Rl+S)-2R(K\bar{x}_{K,l}-l).
	\end{aligned}
	\end{equation}
\end{lemma}
\begin{proof}
	Recall the definition of $\mathcal{L}(K,l,\mu)$ 
	\begin{align*}
	\mathcal{L}(K,l,\mu)
	&= \lim\limits_{T\rightarrow \infty} \frac{1}{T} \mathbb{E}  \sum_{t=0}^{T-1} (x_{t}^{\top} Q_{\mu} x_{t}+ 2x_{t}^{\top}S +u_{t}^{\top} R u_{t}- \mu \bar{\rho}) \\
	& = \mathbb{E}_\tau (x_{t}^{\top} Q_{\mu} x_{t}+ 2x_{t}^{\top}S +(-K x_t + l)^{\top} R (-K x_t + l)- \mu \bar{\rho}),
	\end{align*}
	where $\tau$ denotes the stationary distribution of the steady state. 
	
	Combining the definition of $\bar{x}_{K,l}$ and $\Sigma_{K}$, we obtain
	\begin{equation}\label{equ:lagr}
		\mathcal{L}(K,l,\mu) = \mathrm{tr}\{Q_K(\Sigma_{K} + \bar{x}_{K,l}\bar{x}_{K,l}^{\top})\} + (2S^{\top} -2l^{\top}RK)\bar{x}_{K,l} + l^{\top}Rl- \mu \bar{\rho}.
	\end{equation}
	Then, we derive the gradient with respect to $K$ and $l$, respectively. It follows from (\ref{equ:lagr}) that 
	\begin{align*}
	\nabla_{K} \mathcal{L} &= 2RK(\Sigma_{K} + \bar{x}_{K,l}\bar{x}_{K,l}^\top) + \nabla_{K} \operatorname{tr}\{Q_K(\Sigma_{K} + \bar{x}_{K,l}\bar{x}_{K,l}^\top)\}|_{Q_K = Q_{\mu}+ K^{\top}RK} + \nabla_{K}\{(2S^{\top} - 2l^{\top}RK)\bar{x}_{K,l}\}\\
	&= 2RK(\Sigma_{K} + \bar{x}_{K,l}\bar{x}_{K,l}^\top) -2B^{\top}P_K(A-BK)\Sigma_{K} - 2B^{\top}V^{\top}( Q_{\mu}+ K^{\top}RK)\bar{x}_{K,l}\bar{x}_{K,l}^\top\\
	&~~~~-2Rl\bar{x}_{K,l} - 2B^{\top} V^{\top}(S - K^{\top}Rl)\bar{x}_{K,l}^{\top}\\
	&= 2E_K\Sigma_{K} + 2R(K\bar{x}_{K,l} - l)\bar{x}_{K,l}^{\top} - 2B^{\top}V^{\top}(( Q_{\mu}+ K^{\top}RK)\bar{x}_{K,l} -K^{\top}Rl+S)\bar{x}_{K,l}^{\top},
	\end{align*}
	and
	\begin{align*}
	\nabla_l \mathcal{L}(K,l,\mu) & = \nabla_l \operatorname{tr} \{( Q_{\mu}+ K^{\top}RK)\bar{x}_{K,l}\bar{x}_{K,l}^\top\} + \nabla_{l}\{(2S^{\top} - 2l^{\top}RK)\bar{x}_{K,l}\} + 2Rl \\
	& = 2B^{\top}V^{\top}( Q_{\mu}+ K^{\top}RK)\bar{x}_{K,l}  + 2B^{\top}V^{\top}(S - K^{\top}Rl) -2RK(\bar{x}_{K,l} - l).
	\end{align*}
	Hence, it can be easily observed that 
	$$
	\nabla_{K} \mathcal{L}(K,l,\mu) = 2E_K\Sigma_{K} - \nabla_{l}  \mathcal{L}(K,l,\mu) \bar{x}_{K,l}^{\top}.
	$$
\end{proof}

In view of Proposition \ref{prop:lag}, $\mathcal{L}(K,l,\mu)$ is a quadratic function of $l$, thus $\nabla_{l} \mathcal{L}(K,l,\mu)$ can be derived in another way, i.e.,
\begin{align*}
\nabla_{l} \mathcal{L}(K,l,\mu) & = \nabla_{l} \mathrm{tr}\{P_K(W + (Bl+\bar{w})(Bl+\bar{w})^{\top})\} + \nabla_{l}\{g_{K,l}^{\top}(Bl+\bar{w})\} + 2Rl \\
& = 2B^{\top}P_K(Bl+\bar{w}) + (2B^{\top}PA- 2(B^{\top}PB+R)K)\bar{x}_{K,l} + B^{\top}g_{K,l} \\
& = 2G_{K,l} - 2E_K\bar{x}_{K,l}.
\end{align*}
Hence, it follows that
\begin{align*}
\nabla_{K} \mathcal{L}(K,l,\mu) & = 2E_K\Sigma_{K} - \nabla_{l}  \mathcal{L}(K,l,\mu) \bar{x}_{K,l}^{\top} \\
& = 2E_K(\Sigma_{K} + \bar{x}_{K,l}\bar{x}_{K,l}^{\top}) - 2G_{K,l}\bar{x}_{K,l}^{\top}.
\end{align*}
Noting that 
$
\nabla_{X} \mathcal{L}(X,\mu) =
\left[
\nabla_{K} \mathcal{L} ~~
\nabla_{l} \mathcal{L}
\right],
$ the proof is completed.

\section{Proof of Lemma \ref{lem:gradient dominance}}
As in \citet{fazel2018global}, the gradient dominance property can be derived by analysing the advantage function. We begin by a definition.
\begin{definition}\label{def:def}
	Define the T-truncated value function 
	\begin{equation}\notag
	V_{K,l}^{T}(x)= \mathbb{E} \sum_{t = 0}^{T-1} [c_\mu(x_{t}, u_{t}) -\mathcal{L}(K,l,\mu)| x_{0}=x],
	\end{equation}
	and the T-truncated action-dependent value function 
	\begin{equation}
	Q_{K,l}^{T}(x, u)= c_{\mu}(x,u)-\mathcal{L}(K,l,\mu) + \mathbb{E}V_{K,l}^{T-1}(Ax+Bu+w).
	\end{equation}\notag
	The T-truncated advantage value function is given by
	\begin{equation}\notag
	A_{K,l}^{T}(x, u)= Q_{K,l}^{T}(x, u)-V_{K,l}^{T}(x).
	\end{equation}
\end{definition}
With the T-truncated value function description, we are able to compute the Lagrangian cost difference of two policies $(K,l)$ and $(K',l')$.
\begin{lemma}
	Suppose that both $(K,l)$ and $(K',l')$ are stabilizing. Let $\{x_t'\}$ and $\{u_t'\}$ be sequences generated by following $(K',l')$. Then 
	\begin{equation}\notag
	\mathcal{L}(K',l',\mu)-\mathcal{L}(K,l,\mu)=
	\lim\limits_{T\rightarrow \infty} \frac{1}{T}\sum_{t=0}^{T-1}A_{K,l}^{T}(x_t', u_t').
	\end{equation}
\end{lemma}
\begin{proof}
	It follows from the definition of  $V_{K,l}^{T}(x)$ that
	\begin{align*}
	\mathcal{L}(K',l',\mu)-\mathcal{L}(K,l,\mu)&=
	\lim\limits_{T\rightarrow \infty} \frac{1}{T} \mathbb{E} \sum_{t=0}^{T-1}(c_{\mu}(x_t',u_t')-c_{\mu}(x_t,u_t))\\
	&=\lim\limits_{T\rightarrow \infty}\frac{1}{T} \mathbb{E}\big[\sum_{t=0}^{T-1}(c_{\mu}(x_t',u_t')) - V_{K,l}^{T}(x_t') - T\cdot \mathcal{L}(K,l,\mu)\big]\\
	& = \lim\limits_{T\rightarrow \infty}\frac{1}{T} \mathbb{E}\big[\sum_{t=0}^{T-1}\big(c_{\mu}(x_t',u_t')+V_{K,l}^{T}(x_t')-V_{K,l}^{T}(x_t')\big) - V_{K,l}^{T}(x_t') - T\cdot \mathcal{L}(K,l,\mu)\big]\\
	& = \lim\limits_{T\rightarrow \infty} \frac{1}{T} \mathbb{E}\sum_{t=0}^{T-1}(c_{\mu}(x_t',u_t')-\mathcal{L}(K,l,\mu)+ V_{K,l}^{T}(x_{t+1}')-V_{K,l}^{T}(x_t')) \\
	& = \lim\limits_{T\rightarrow \infty} \frac{1}{T} \mathbb{E}\sum_{t=0}^{T-1}(Q_{K,l}^{T}(x_t', u_t')-V_{K,l}^{T}(x_t'))\\
	& = \lim\limits_{T\rightarrow \infty} \frac{1}{T} \mathbb{E}\sum_{t=0}^{T-1}A_{K,l}^{T}(x_t', u_t')
	\end{align*}
\end{proof}

Then, we show that the advantage value function has a closed-form.
\begin{lemma}
	The advantage value function $A_{K,l}(x, u) = \lim\limits_{T\rightarrow \infty} A_{K,l}^{T}(x,u)$ under policy $(K',l')$ is given as
	\begin{align*}
	A_{K,l}(x,u) &= Q_{K,l}(x,u) - V_{K,l}(x)\\
	&= \lim\limits_{T\rightarrow \infty} Q_{K,l}^{\top}(x,u) - V_{K,l}^{\top}(x) \\
	&=2 x^{\top}\left(K'-K\right)^{\top} E_{K} x+x^{\top}\left(K'-K\right)^{\top}\left(R+B^{\top} P_{K} B\right)\left(K'-K\right) x \\
	&~~~~-2G_{K,l}^{\top}(K'-K)x - 2(l'-l)^{\top}(R+B^{\top}P_{K} B)(K'-K)x\\
	&~~~~-2(l'-l)^{\top}E_Kx+2(l'-l)^{\top}G_{K,l}+(l'-l)^{\top}(R+B^{\top}P_{K} B)(l'-l)
	\end{align*}		
\end{lemma}
\begin{proof}
	By Definition \ref{def:def}, it follows that
	\begin{align*}
	A_{K,l}(x,u) &= Q_{K,l}(x,u) - V_{K,l}(x)\\
	&= \lim\limits_{T\rightarrow \infty} Q_{K,l}^{\top}(x,u) - V_{K,l}^{\top}(x) \\
	& = x^{\top}Q_{\mu}x + 2S^{\top}x + (-K'x+l')^{\top}R(-K'x+l') -\mathcal{L}(K,l,\mu) \\
	&~~~~+ \mathbb{E}_wV_{K,l}\big((A-BK')x+Bl'+w\big) - V_{K,l}(x)\\
	&= x^{\top}(Q_{\mu}+(K')^{\top} R K') x+x^{\top}(A-B K')^{\top} P_{K}(A-B K') \\
	&~~~~+(2S^{\top} -2(l')^{\top}RK')x + 2(Bl'+\bar{w})^{\top}P_K(A-BK')x + g_{K,l}^{\top}(A-BK')x\\
	&~~~~+ tr\{P_K(W + (Bl'+\bar{w})(Bl'+\bar{w})^{\top} )\} + g_{K,l}^{\top}(Bl'+\bar{w}) + (l')^{\top}Rl' - \mathcal{L}(K,l,\mu) -V_{K,l}(x)\\
	&= x^{\top}(Q_{\mu}+(K')^{\top} R K') x+x^{\top}(A-B K')^{\top} P_{K}(A-B K') \\
	&~~~~ -x^{\top}(Q_{\mu}+(K)^{\top} R K) x-x^{\top}(A-B K)^{\top} P_{K}(A-B K)	\\
	&~~~~ + (2S^{\top} -2(l')^{\top}RK')x + 2(Bl'+\bar{w})^{\top}P_K(A-BK')x + g_{K,l}^{\top}(A-BK')x \\
	&~~~~ - (2S^{\top} -2l^{\top}RK)x - 2(Bl+\bar{w})^{\top}P_K(A-BK)x - g_{K,l}^{\top}(A-BK)x\\
	&~~~~ + tr\{P_K(W + (Bl'+\bar{w})(Bl'+\bar{w})^{\top} )\} + g_{K,l}^{\top}(Bl'+\bar{w}) + (l')^{\top}Rl'\\
	&~~~~ - tr\{P_K(W + (Bl+\bar{w})(Bl+\bar{w})^{\top} )\} - g_{K,l}^{\top}(Bl+\bar{w}) - l^{\top}Rl
	\end{align*}		
	By reorganizing the terms and using the definition of $G_{K,l}$, the proof is completed.
\end{proof}
We are now ready to establish the key lemma that leads to the local gradient dominance.
\begin{lemma}  For a stabilizing policy $X \in \mathcal{S}$, it holds that
	\begin{equation}\notag
	\mathcal{L}(K,l,\mu)-\mathcal{L}(K^*,l^*,\mu)
	\leq \frac{\|\Phi^*\|}{4\sigma_{min}(R)\sigma_{min}(\Phi_{K,l})^2}\operatorname{tr} \left\{
	\left[
	\nabla_{K} \mathcal{L} ~~
	\nabla_{l} \mathcal{L}
	\right]^{\top}	
	\left[
	\nabla_{K} \mathcal{L} ~~
	\nabla_{l} \mathcal{L}
	\right]\right\}
	\end{equation}
\end{lemma}
\begin{proof}
Note that $A_{K,l}(x,u)$ can be further reorganized as
\begin{equation}\notag
\begin{aligned}
A_{K,l}(x,u) &= [(K'-K)x-(l'-l) + (R+B^{\top}P_KB)^{-1}(E_Kx-G_{K,l})]^{\top}(R+B^{\top}P_KB)\\
&~~~~\times [(K'-K)x-(l'-l) + (R+B^{\top}P_KB)^{-1}(E_Kx-G_{K,l})]\\
&~~~~- (E_Kx-G_{K,l})^{\top}(R+B^{\top}P_KB)^{-1}(E_Kx-G_{K,l}) \\
& \geq - (E_Kx-G_{K,l})^{\top}(R+B^{\top}P_KB)^{-1}(E_Kx-G_{K,l}).
\end{aligned}		
\end{equation}

Let $\{x_t^*\}$ and $\{u_t^*\}$ be sequences generated by following the optimal policy $(K^*,l^*)$ and $\Phi^*$ be the correlation matrix $\Phi_{K^*,l^*}$. Then, it follows that
\begin{equation}\notag
\begin{aligned}
&\mathcal{L}(K,l,\mu)-\mathcal{L}(K^*,l^*,\mu)\\
&=-
\lim\limits_{T\rightarrow \infty} \frac{1}{T}\mathbb{E}\sum_{t=0}^{T-1}A_{K,l}^{T}(x_t^*, u_t^*)\\
& \leq \lim\limits_{T\rightarrow \infty} \frac{1}{T}\mathbb{E}\sum_{t=0}^{T-1} \operatorname{tr}\left\{(E_Kx_t^*-G_{K,l})^{\top}(R+B^{\top}P_KB)^{-1}(E_Kx_t^*-G_{K,l})\right\} \\
& = \lim\limits_{T\rightarrow \infty} \frac{1}{T}\mathbb{E}\sum_{t=0}^{T-1} \operatorname{tr} \left\{ 		
\left[\begin{array}{c}
x_t^* \\
-1
\end{array}\right]
\left[\begin{array}{c}
x_t^* \\
-1
\end{array}\right]^{\top}[E_K~~G_{K,l}]^{\top}(R+B^{\top}P_KB)^{-1}[E_K~~G_{K,l}]
\right\}\\
&= \operatorname{tr} \left\{ \left(\lim\limits_{T\rightarrow \infty} \frac{1}{T}\mathbb{E}\sum_{t=0}^{T-1} 		
\left[\begin{array}{c}
x_t^* \\
-1
\end{array}\right]
\left[\begin{array}{c}
x_t^* \\
-1
\end{array}\right]^{\top}\right) [E_K~~G_{K,l}]^{\top}(R+B^{\top}P_KB)^{-1}[E_K~~G_{K,l}]
\right\}\\
& = \operatorname{tr} \left\{ \Phi^* [E_K~~G_{K,l}]^{\top}(R+B^{\top}P_KB)^{-1}[E_K~~G_{K,l}]
\right\}\\
& \leq \|\Phi^*\| \operatorname{tr} \left\{ [E_K~~G_{K,l}]^{\top}(R+B^{\top}P_KB)^{-1}[E_K~~G_{K,l}]
\right\}\\
& = \|\Phi^*\| \operatorname{tr} \left\{ (R+B^{\top}P_KB)^{-1}[E_K~~G_{K,l}][E_K~~G_{K,l}]^{\top} \right\}\\
& \leq \|\Phi^*\| \|(R+B^{\top}P_KB)^{-1}\|\operatorname{tr} \left\{ [E_K~~G_{K,l}][E_K~~G_{K,l}]^{\top} \right\}\\
& \leq \frac{\|\Phi^*\|}{\sigma_{min}(R)}
\operatorname{tr} \left\{ [E_K~~G_{K,l}][E_K~~G_{K,l}]^{\top} \right\}\\
\end{aligned}
\end{equation}

To make the connections between the policy gradients and the cost difference clear, recall that
\begin{equation}\notag
\begin{aligned}
&\operatorname{tr} \left\{
\left[
\nabla_{K} \mathcal{L} ~~
\nabla_{l} \mathcal{L}
\right]^{\top}
\left[
\nabla_{K} \mathcal{L} ~~
\nabla_{l} \mathcal{L}
\right]\right\}  = 4\operatorname{tr}\left\{\Phi_{K,l}\Phi_{K,l}^{\top}[E_K~~G_{K,l}]^{\top}[E_K~~G_{K,l}]
\right\}
\end{aligned}
\end{equation}
Thus, the gap $\mathcal{L}(K,l,\mu)-\mathcal{L}(K^*,l^*,\mu)$ can be further bounded by the gradient, i.e.,
\begin{equation}\notag
\begin{aligned}
&\mathcal{L}(K,l,\mu)-\mathcal{L}(K^*,l^*,\mu)\\
& \leq\frac{\|\Phi^*\|}{\sigma_{min}(R)} \operatorname{tr} \left\{ [E_K~~G_{K,l}][E_K~~G_{K,l}]^{\top} \right\}\\
&=\frac{\|\Phi^*\|}{4\sigma_{min}(R)} \operatorname{tr} \left\{(\Phi_{K,l}\Phi_{K,l}^{\top} )^{-1}
\operatorname{tr} \left\{
\left[
\nabla_{K} \mathcal{L} ~~
\nabla_{l} \mathcal{L}
\right]^{\top}
\left[
\nabla_{K} \mathcal{L} ~~
\nabla_{l} \mathcal{L}
\right]\right\}\right\} \\
& \leq \frac{\|\Phi^*\|}{4\sigma_{min}(R)\sigma_{min}(\Phi_{K,l})^2}\operatorname{tr} \left\{
\left[
\nabla_{K} \mathcal{L} ~~
\nabla_{l} \mathcal{L}
\right]^{\top}
\left[
\nabla_{K} \mathcal{L} ~~
\nabla_{l} \mathcal{L}
\right]\right\} \\
\end{aligned}
\end{equation}
\end{proof}

Since the lower bound of $\sigma_{min}(\Phi_{K,l})$ is zero, we cannot find a uniform gradient dominance constant. However, we notice that the positive definite correlation matrix $\Phi_{K,l}$ (\ref{def:phi}) is continuous with respect to $K$ and $l$. Thus, if we restrict the policy $X$ in a compact set, its minimal eigenvalue must be positive. The proof is completed by considering the compact sub-level set in (\ref{def:sublevel}).

\section{Proof of Lemma \ref{lem:lip}}
In this section, we show the Lipschitz continuity for the Lagrangian function and its gradient, respectively.
\subsection{Proof of Lipschitz continuity of the Lagrangian Function}
We introduce a auxiliary cost function to facilitate the analysis.
\begin{lemma}
	Define $C(K) =  \mathrm{tr}(P_KW)$, then it follows that
	\begin{equation}\notag
	C(K) = \mathrm{tr}\bigl[(Q_{\mu} + K^TRK)\Sigma_{K}\bigr] \leq \mathcal{L}(K,l) + S^TQ_{\mu}^{-1}S.
	\end{equation}
\end{lemma}

\begin{proof}
	Comparing the definition of $\mathcal{L}(K,l,\mu)$ and $C(K)$, we obtain that
	\begin{equation}\notag
	\begin{aligned}
	\mathcal{L}(K,l,\mu) &= \mathrm{tr}\bigl[(Q_{\mu} + K^TRK)(\Sigma_{K} + \bar{x}_{K,l}\bar{x}_{K,l}^{\top})\bigr] + (2S^{\top} -2l^{\top}RK)\bar{x}_{K,l} + l^{\top}Rl\\
	&=\mathrm{tr}\bigl[(Q_{\mu} + K^TRK)\Sigma_{K}\bigr] + (Q_{\mu}\bar{x}_{K,l} +S)^TQ_{\mu}^{-1}(Q_{\mu}\bar{x}_{K,l} +S) - S^TQ_{\mu}^{-1}S \\
	&~~~~+(K\bar{x}_{K,l} -l)^TR(K\bar{x}_{K,l} -l) \\
	& \geq C(K) - S^TQ_{\mu}^{-1}S
	\end{aligned}
	\end{equation}
\end{proof}
In fact, $C(K)$ is the cost function of the classical LQR formulation, see \citet{fazel2018global, bu2019lqr}. By introducing it, we build a connection between $\mathcal{L}(K,l,\mu)$ and $C(K)$. Thus, some results in~\citet{fazel2018global} can be utilized for our analysis. 

We now present some technical lemmas.

\begin{lemma}\label{lem:fazel}
	Suppose that $K$ is stabilizing, i.e., $\rho(A-BK) < 1$. Then we have following results.
	(a)
	\begin{equation}\notag
	\left\|\Sigma_{K}\right\| \leq \frac{ C(K)}{\sigma_{\min }(Q_{\mu})},~~~~ \|P_K\| \leq \frac{C(K)}{\sigma_{\min }(W)}.
	\end{equation}
	(b)
	If
	\begin{equation}\notag
	\left\|K^{\prime}-K\right\| \leq \min \left(\frac{\sigma_{\min }(Q_{\mu}) \sigma_{\min }(W)}{4 C(K)\|B\|(\|A-B K\|+1)},\|K\|\right),
	\end{equation}	
	then it follows that
	\begin{equation}\notag
	\|P_{K'} - P_K \| \leq  6\|K\|\|R\| \left(\frac{C(K)}{\sigma_{\min }(Q_{\mu}) \sigma_{\min }(W)}\right)^{2}(\|K\|\|B\|\|A-B K\|+\|K\|\|B\|+1)\|K-K^{\prime}\|.
	\end{equation}
	(c) The norm $\|K\|$ can be bounded by
	\begin{equation}\notag
	\|K\| \leq \frac{1}{\sigma_{\min }(R)}\left(\sqrt{\frac{\left\|R+B^{\top} P_{K} B\right\|\left(C(K)-C\left(K^{*}\right)\right)}{\mu}}+\|B^{\top} P_{K} A\|\right)
	\end{equation}
	(d) \begin{equation}\notag
	\operatorname{tr}\left(\Sigma_{K}\right) \geq \frac{\sigma_{min}(W)}{2\left(1-\rho\left(A-B K\right)\right)}.
	\end{equation}
	As a consequence, $\frac{1}{1-\rho(A-BK)}$ is bounded, i.e.,
	\begin{equation}\notag
	\frac{1}{1-\rho(A-BK)} \leq \frac{2n\|\Sigma_{K}\|}{\sigma_{min}(W)} \leq \frac{2nC(K)}{\sigma_{min}(W) \sigma_{min}(Q_{\mu})}.
	\end{equation}
\end{lemma}

\begin{proof}
	The proof follows the results of \citet{fazel2018global}.
\end{proof}

\begin{lemma}\label{lem:tec_V}
	Define $V_K = (I - (A-BK))^{-1}$. Suppose that
	$
	\|K' - K\| \leq \frac{1-\rho(A-BK)}{2\|B\|},
	$
	it follows that
	\begin{equation}\notag
	\|V_{K'} - V_{K}\| \leq \frac{2 \|B\| \|K' - K\|}{1-\rho(A-BK)}
	\end{equation}
\end{lemma}

\begin{proof}
	The proof follows immediately from matrix inverse perturbation theorem~\citep{horn2012matrix}.
\end{proof}

\begin{lemma} \label{lem:EK}
	$\mathrm{tr}(E_K^TE_K)$ can be bounded as
	\begin{equation}
	\mathrm{tr}(E_K^TE_K) \leq \operatorname{tr} \left\{ [E_K~~G_{K,l}]^T[E_K~~G_{K,l}] \right\} \leq \frac{\|R+B^TP_KB\|}{\phi_{\alpha}} (\mathcal{L}(K,l)- \mathcal{L}(K^*,l^*)),
	\end{equation}
	where $\phi_{\alpha}$ is defined in Lemma \ref{lem:gradient dominance}.
\end{lemma}

\begin{proof}
	Let $X^{\prime} =X-\left(R+B^{\top} {P}_{K} B\right)^{-1} \left[E_K~~G_{K,l}\right]$, we have
	\begin{equation}\notag
	\begin{aligned}
	\mathcal{L}(K,l)- \mathcal{L}(K^*,l^*) &\geq \mathcal{L}(K,l)- \mathcal{L}(K',l')\\
	&= \operatorname{tr} \left\{ [E_K~~G_{K,l}]^T\left(R+B^{\top} {P}_{K} B\right)^{-1}[E_K~~G_{K,l}] \Phi_{K',l'}\right\}\\
	& \geq \frac{\sigma_{min}(\Phi_{K',l'})}{\|R+B^TP_KB\|}\operatorname{tr} \left\{ [E_K~~G_{K,l}]^T[E_K~~G_{K,l}] \right\} \\
	& \geq \frac{\phi_{\alpha}}{\|R+B^TP_KB\|}\operatorname{tr} \left\{ [E_K~~G_{K,l}]^T[E_K~~G_{K,l}] \right\}.
	\end{aligned}
	\end{equation}
	The last inequality is obtained by choosing a sufficiently large $\alpha$.
\end{proof}

\begin{lemma}
	For all $K'$ such that
	\begin{equation}\notag
	\left\|K^{\prime}-K\right\| \leq \min \left(\frac{\sigma_{\min }(Q_{\mu}) \sigma_{\min }(W)}{4 C(K)\|B\|(\|A-B K\|+1)},\|K\|,\frac{1-\rho(A-BK)}{2\|B\|}\right),
	\end{equation}	
	it follows that,
	\begin{equation}\notag
	\|	g_{K',l}^T-g_{K,l}^T\| \leq c_{K1} \|K'-K\|.
	\end{equation}
\end{lemma}

\begin{proof}
	It follows from the definition of $g_{K,l}$ that
	\begin{equation}\notag
	\begin{aligned}
	g_{K',l}^T-g_{K,l}^T &= 2[-l^{\top}E_{K'}+S^{\top} + \bar{w}^{\top}P_{K'}(A-BK')]V_{K'} - \\
	&~~~~ 2[-l^{\top}E_{K}+S^{\top} + \bar{w}^{\top}P_{K}(A-BK)]V_{K}\\
	& = 2[-l^{\top}(E_{K'}-E_K) + \bar{w}^{\top}P_{K'}(A-BK') -\bar{w}^{\top}P_{K}(A-BK)]V_{K'}\\
	& ~~~~ + 2[-l^{\top}E_{K}+S^{\top} + \bar{w}^{\top}P_{K}(A-BK)](V_{K'}-V_{K})
	\end{aligned}
	\end{equation}
	By lemma \ref{lem:tec_V}, we have
	\begin{equation}\notag
	\|V_{K'} - V_K\| \leq  \frac{2 \|B\| \|K' - K\|}{1-\rho(A-BK)}.
	\end{equation}
	Further,
	\begin{equation}\notag
	\|V_{K'}\| = \|V_{K'} - V_K + V_K\| \leq \|V_{K'} - V_K\| + \|V_K\|=\frac{1+2 \|B\| \|K' - K\|}{1-\rho(A-BK)}.
	\end{equation}
	By lemma \ref{lem:EK},
	\begin{equation}\notag
	\|E_K\| \leq \sqrt{\frac{\|R+B^TP_KB\|}{\phi_{\alpha}} (\mathcal{L}(K,l)- \mathcal{L}(K^*,l^*))}.
	\end{equation}
	Also note that
	\begin{equation}\notag
	\|P_{K'}(A-BK') - P_{K}(A-BK)\| \leq \|P_{K'}-P_K\|\|A-BK'\| +\| P_K\|\|B\|K'-K\|.
	\end{equation}
	By using the assumptions on $\|K'-K\|$ and the bounds in lemma \ref{lem:fazel},
	we finally obtain that
	\begin{equation}\notag
	\|	g_{K',l}^T-g_{K,l}^T\| \leq c_1(K,l) \|K'-K\|,
	\end{equation}
	where $c_1(K,l)$ is polynomial in $C(K), \|A\|, \|B\|, \frac{1}{\sigma_{min}(W)}, \frac{1}{\sigma_{min}(Q_{\mu})}, \frac{1}{\sigma_{min}(R)}, \|l\|, \|\bar{w}\|$.
\end{proof}

Equipped with the above lemmas, we are now ready to find the Lipschitz constants of $\mathcal{L}(K,l,\mu)$ with respect to $K$ and $l$, respectively.
\begin{lemma} \label{lem:K}
	Suppose that
	\begin{equation}\notag
	\left\|K^{\prime}-K\right\| \leq \min \left(\frac{\sigma_{\min }(Q_{\mu}) \sigma_{\min }(W)}{4 C(K)\|B\|(\|A-B K\|+1)}, \|K\|, \frac{1-\rho(A-BK)}{2\|B\|}\right),
	\end{equation}
	then
	\begin{equation}\notag
	\|\mathcal{L}(K',l) - \mathcal{L}(K,l)\| \leq  c_2(K,l) \|K'-K\|
	\end{equation}
\end{lemma}

\begin{proof}
	Note that
	\begin{equation}\notag
	\mathcal{L}(K',l) - \mathcal{L}(K,l) = \mathrm{tr}\bigl[(P_{K'}-P_K)(W + (Bl+\bar{w})(Bl+\bar{w})^{\top} )\bigr] + (g_{K',l}-g_{K,l})^{\top}(Bl+\bar{w}).
	\end{equation}
	Using the bounds built in above lemmas, we can find $c_2(K,l)$ as a polynomial in $C(K), \|A\|, \|B\|$, $\frac{1}{\sigma_{min}(W)}, \frac{1}{\sigma_{min}(Q_{\mu})}, \frac{1}{\sigma_{min}(R)}, \|l\|, \|\bar{w}\|, \|W\|$.
\end{proof}

\begin{lemma} \label{lem:l}
	Suppose that
	$
	\|l'-l\| \leq \|l\| + \bar{w},
	$
	then
	\begin{equation}\notag
	\|\mathcal{L}(K,l') - \mathcal{L}(K,l)\| \leq  c_3(K,l) \|l'-l\|.
	\end{equation}
\end{lemma}

\begin{proof}
	Since $\mathcal{L}(K,l)$ is quadratic in $l$, the analysis is much simpler. We have
	\begin{equation}\notag
	\begin{aligned}
	\mathcal{L}(K,l') - \mathcal{L}(K,l) &=l'^T(R+B^TP_KB)l' -l^T(R+B^TP_KB)l + 2\bar{w}^TP_KB(l'-l)\\
	&~~~~ + g_{l'}^T(Bl'+\bar{w}) -g_{l}^T(Bl+\bar{w})\\
	&= (l'-l)^T(R+B^TP_KB)l' + l^T(R+B^TP_KB)(l'-l) \\ &~~~~+2\bar{w}^TP_KB(l'-l) + g_{l'}^TB(l'-l) + (g_{l'}^T-g_l^T)(Bl+\bar{w}).
	\end{aligned}
	\end{equation}
	Since
	\begin{equation}\notag
	\|g_{l'}^T-g_l^T\| \leq 2 \| l'-l\|\|E_K\|\|V_K\|,
	\end{equation}
	it follows that
	\begin{equation}\notag
	\|g_{l'} \| \leq \|g_l\| + \|g_{l'}^T-g_l^T\| \leq
	2 (2\|l\|+\|\bar{w}\|)\|E_K\|\|V_K\| + 2 \| -l^{\top}E_K+S^{\top} + \bar{w}^{\top}P_K(A-BK) \|\|V_K\|.
	\end{equation}
	Also using the technical lemmas, we obtain 
	\begin{equation}\notag
	\|\mathcal{L}(K,l') - \mathcal{L}(K,l)\| \leq  c_3(K,l) \|l'-l\|,
	\end{equation}
	where $c_3(K,l)$ is polynomial in $C(K), \|R\|, \|A\|, \|B\|, \frac{1}{\sigma_{min}(W)}, \frac{1}{\sigma_{min}(Q_{\mu})}, \frac{1}{\sigma_{min}(R)}, \|l\|, \|\bar{w}\|$.
\end{proof}

Combining Lemma \ref{lem:K} and Lemma \ref{lem:l}, the Lipschitz constant can be found.
\begin{lemma}[The cost is locally Lipschitz.]
	There exist positive scalars $(L_1, \gamma_{1})$ that depends on the current policy $X = [K~~l]$, such that for all policies $X'$ satisfying $\| X'-X\| \leq \gamma_{1}$, the cost difference is Lipschitz bounded, namely,
	\begin{equation}
	\|\mathcal{L}(X',\mu) - \mathcal{L}(X,\mu)\| \leq L_1 \| X'-X\|.
	\end{equation}
\end{lemma}

\begin{proof}
	By choosing $\gamma_{1}$ such that the assumptions in Lemma \ref{lem:K} and Lemma \ref{lem:l} hold for all $X' \in \{X'|\| X'-X\| \leq \gamma_{1}\}$, if follows that
	\begin{equation}
	\begin{aligned}
	\|\mathcal{L}(K',l') - \mathcal{L}(K,l)\| &= \|\mathcal{L}(K',l') -\mathcal{L}(K',l) + \mathcal{L}(K',l)
	- \mathcal{L}(K,l)\| \\
	&\leq \|\mathcal{L}(K',l') -\mathcal{L}(K',l)\| + \| \mathcal{L}(K',l)
	- \mathcal{L}(K,l)\|\\
	& \leq c_3(K',l) \|l'-l\| + c_2(K,l) \|K'-K\| \\
	&\leq L_1 \|X'-X\|.
	\end{aligned}
	\end{equation}
\end{proof}
\subsection{Proof of the Lipschitz Continuity of the Gradient}
Analogy to the above derivation, we first establish the Lipschitz property for the gradient $\nabla_{K}\mathcal{L}$ and $\nabla_{l}\mathcal{L}$ and then combine them.
\begin{lemma}
	Suppose that
	\begin{equation}\notag
	\left\|K^{\prime}-K\right\| \leq \min \left(\frac{\sigma_{\min }(Q_{\mu}) \sigma_{\min }(W)}{4 C(K)\|B\|(\|A-B K\|+1)}, \|K\|, \frac{1-\rho(A-BK)}{2\|B\|}\right),
	\end{equation}
	then it follows that
	\begin{equation}\notag
	\|\Phi_{K',l} - \Phi_{K,l}\| \leq c_4(K,l) \|K'-K\|.
	\end{equation}
\end{lemma}

\begin{proof}
	Note that
	\begin{equation}\label{equ:second}
	\begin{aligned}
	\|\Phi_{K',l} - \Phi_{K,l}\| &\leq \operatorname{tr} (\Phi_{K',l} - \Phi_{K,l})\\
	&\leq n \| \Sigma_{K'}-\Sigma_{K}\| +  \|\bar{x}_{K'}\bar{x}_{K'}^T -  \bar{x}_{K}\bar{x}_{K}^T\|.
	\end{aligned}
	\end{equation}
	It has been shown in \citet{fazel2018global} that
	\begin{equation}\notag
	\left\|\Sigma_{K^{\prime}}-\Sigma_{K}\right\| \leq 4\left(\frac{C(K)}{\sigma_{\min }(Q_{\mu})}\right)^{2} \frac{\|B\|(\|A-B K\|+1)}{\sigma_{min}(W)}\left\|K-K^{\prime}\right\|.
	\end{equation}
	For the second term in (\ref{equ:second})
	\begin{equation}\notag
	\|\bar{x}_{K'}\bar{x}_{K'}^T -  \bar{x}_{K}\bar{x}_{K}^T\| = \|(\bar{x}_{K'}- \bar{x}_{K}) \bar{x}_{K'} +  \bar{x}_{K}(\bar{x}_{K'}- \bar{x}_{K})\|,
	\end{equation}
	we have
	\begin{equation}\notag
	\|\bar{x}_{K'}- \bar{x}_{K}\| = \|(V_{K'} - V_K)(Bl+\bar{w}) \| \leq \frac{2 \|B\|\|Bl+\bar{w}\| \|K' - K\|}{1-\rho(A-BK)}
	\end{equation}
	and
	\begin{equation}\notag
	\begin{aligned}
	\|\bar{x}_{K'}\| &\leq \|\bar{x}_{K} \| + \| \bar{x}_{K'}- \bar{x}_{K}\| \\
	&\leq \frac{(2 \|B\| \|K' - K\|+1)\|Bl+\bar{w}\|}{1-\rho(A-BK)}.
	\end{aligned}
	\end{equation}
	Thus,
	\begin{equation}\label{equ:hh}
	\begin{aligned}
	\|\bar{x}_{K'}\bar{x}_{K'}^T -  \bar{x}_{K}\bar{x}_{K}^T\| &\leq 2\left(\frac{\|Bl+\bar{w}\|}{1-\rho(A-BK)}\right)^2\|B\|(2+2\|B\|\|K'-K\|) \|K'-K\|\\
	& \leq 6\left(\frac{\|Bl+\bar{w}\|}{1-\rho(A-BK)}\right)^2\|B\| \|K'-K\|
	\end{aligned}
	\end{equation}
	Combining (\ref{equ:second}) and (\ref{equ:hh}), we can find $c_4(K,l)$ as a polynomial in 
	$C(K), \|R\|, \|A\|, \|B\|, \frac{1}{\sigma_{min}(W)}$, $~ \frac{1}{\sigma_{min}(Q_{\mu})}, \frac{1}{\sigma_{min}(R)}, \|l\|, \|\bar{w}\|,n$.
\end{proof}

\begin{lemma}
	Suppose that
	\begin{equation}\notag
	\left\|K^{\prime}-K\right\| \leq \min \left(\frac{\sigma_{\min }(Q_{\mu}) \sigma_{\min }(W)}{4 C(K)\|B\|(\|A-B K\|+1)},\|K\|, \frac{1-\rho(A-BK)}{2\|B\|}\right),
	\end{equation}
	it follows that
	\begin{equation}\notag
	\|\nabla_{K',l} \mathcal{L}- \nabla_{K,l} \mathcal{L}\| \leq c_5(K,l) \|K'-K\|.
	\end{equation}
\end{lemma}

\begin{proof}
	Suppose that $l$ is fixed, then we have
	\begin{equation}\notag
	\begin{aligned}
	\|\nabla_{K',l} \mathcal{L}- \nabla_{K,l} \mathcal{L}\| &= 2\|\left[
	E_{K'}~~
	G_{K',l}
	\right] \Phi_{K',l} -\left[
	E_K~~
	G_{K,l}
	\right] \Phi_{K,l}\| \\
	& \leq 2 \| \left[E_{K'}-E_K~~~~G_{K',l}-G_{K,l}\right] \Phi_{K',l}  + \left[E_K~~G_{K,l}\right] (\Phi_{K',l}-\Phi_{K,l}) \|. \\
	\end{aligned}
	\end{equation}
	$\|\Phi_{K,l}\|$ can be bounded as
	\begin{equation}\notag
	\begin{aligned}
	\|\Phi_{K,l}\| & \leq 1 + \operatorname{tr}(\Sigma_K + \bar{x}_{K}\bar{x}_{K}^T) \\
	& \leq 1+ n \|\Sigma_K\| +  \|\bar{x}_{K} \|^2 \\
	& \leq 1 + n\frac{ C(K)}{\sigma_{\min }(Q_{\mu})} + \left(\frac{\|Bl+\bar{w}\|}{1-\rho(A-BK)}\right)^2.
	\end{aligned}
	\end{equation}
	Further, using the conditions on $K$, it follows that
	\begin{equation}\notag
	\begin{aligned}
	\|\Phi_{K',l}\|& \leq \|\Phi_{K,l}\|+\|\Phi_{K',l}-\Phi_{K,l}\| \\
	& \leq 1 + \frac{5nC(K)}{\sigma_{\min }(Q_{\mu})} + \left(\frac{\|Bl+\bar{w}\|}{1-\rho(A-BK)}\right)^2 +  \frac{3n\|Bl+\bar{w}\|^2}{1-\rho(A-BK)}.
	\end{aligned}
	\end{equation}
	Also,
	\begin{equation}\notag
	\|G_{K',l}-G_{K,l}\| = \|B^{\top}(P_{K'} - P_K)Bl + B^{\top}(P_{K'}-P_K)\bar{w} + \frac{1}{2} B^{\top}(g_{K',l}- g_{K,l})\|
	\end{equation}
	can be bounded by $\|K'-K\|$.
	From lemma \ref{lem:EK}, we obtain
	\begin{equation}\notag
	\|[E_K ~~ G_{K,l}]\| \leq \sqrt{\frac{\|R+B^TP_KB\|}{\sigma_{min}(\Phi_{K',l'})} (\mathcal{L}(K,l)- \mathcal{L}(K^*,l^*))}.
	\end{equation}
	Combining the above inequalities, we conclude that
	\begin{equation}\notag
	\|\nabla_{K',l} \mathcal{L}- \nabla_{K,l} \mathcal{L}\| \leq c_5(K,l) \|K'-K\|,
	\end{equation}
	where $c_5(K,l)$  is polynomial in $C(K), \|R\|, \|A\|, \|B\|, \frac{1}{\sigma_{min}(W)}, \frac{1}{\sigma_{min}(Q_{\mu})}, \frac{1}{\sigma_{min}(R)}, \|l\|, \|\bar{w}\|, n$.
\end{proof}

\begin{lemma}
	Suppose that
$
	\|l'-l\| \leq \|l\|,
$
	then we have
	\begin{equation}\notag
	\|\nabla_{K,l'} \mathcal{L}- \nabla_{K,l} \mathcal{L}\| \leq c_6(K,l) \|l'-l\|,
	\end{equation}
	where $c_6(K,l)$  is polynomial in $C(K), \|R\|, \|A\|, \|B\|, \frac{1}{\sigma_{min}(W)}, \frac{1}{\sigma_{min}(Q_{\mu})}, \frac{1}{\sigma_{min}(R)}, \|l\|, \|\bar{w}\|, n$.
\end{lemma}

\begin{proof}
	Note that
	\begin{equation}\notag
	\begin{aligned}
	\|\nabla_{K,l'} \mathcal{L}- \nabla_{K,l} \mathcal{L}\| &= 2\|\left[
	E_{K}~~
	G_{K,l'}
	\right] \Phi_{K,l'} -\left[
	E_K~~
	G_{K,l}
	\right] \Phi_{K,l}\| \\
	& = 2 \| \left[0~~~~G_{K,l'}-G_{K,l}\right] \Phi_{K,l'}  + \left[E_K~~G_{K,l}\right] (\Phi_{K,l'}-\Phi_{K,l}) \|, \\
	\end{aligned}
	\end{equation}
	Then, combining
	\begin{equation}\notag
	\|G_{K,l'}-G_{K,l}\| = \frac{1}{2}\| B\|\|g_{K,l'}- g_{K,l}\| \leq \| B\|\|E_K\|\|V_K\|\| l'-l\|,
	\end{equation}
	and
	\begin{equation}\notag
	\begin{aligned}
	\|\Phi_{K,l'} - \Phi_{K,l}\| &\leq  \operatorname{tr}(\Phi_{K,l'} - \Phi_{K,l})\\
	&=  \|\bar{x}_{l'}\bar{x}_{l'}^T -  \bar{x}_{l}\bar{x}_{l}^T\|\\
	& =  	 \|(\bar{x}_{l'}- \bar{x}_{l}) \bar{x}_{l'} +  \bar{x}_{l}(\bar{x}_{l'}- \bar{x}_{l})\| \\
	& \leq \big[2\|Bl+\bar{w}\|+ \|B\|\|l\|\big]\|B\|\|V_K\|^2\|l'-l\|,
	\end{aligned}
	\end{equation}
the proof is completed.
\end{proof}

\begin{lemma}[The gradient is locally Lipschitz.]
	There exist positive scalars $(L_2, \gamma_{2})$ that depends on the current policy $X = [K~~l]$, such that for all policies $X'$ satisfying $\| X'-X\| \leq \gamma_{2}$, the gradient difference is Lipschitz bounded, namely,
	\begin{equation}\notag
	\|\nabla_{X'}\mathcal{L} - \nabla_{X}\mathcal{L}\| \leq L_2 \| X'-X\|.
	\end{equation}
\end{lemma}

\begin{proof}
	By choosing $\gamma_{2}$ such that the assumptions in Lemma \ref{lem:K} and Lemma \ref{lem:l} hold for all $X' \in \{X'|\| X'-X\| \leq \gamma_{2}\}$, if follows that
	\begin{equation}\notag
	\begin{aligned}
	\|\nabla_{X'}\mathcal{L} - \nabla_{X}\mathcal{L}\| &= \|\nabla_{K',l'}\mathcal{L} -\nabla_{K',l}\mathcal{L}+ \nabla_{K',l}\mathcal{L}
	- \nabla_{K,l}\mathcal{L}\| \\
	&\leq \|\nabla_{K',l'}\mathcal{L} -\nabla_{K',l}\mathcal{L}\| + \| \nabla_{K',l}\mathcal{L}
	- \nabla_{K,l}\mathcal{L}\|\\
	& \leq c_6(K',l) \|l'-l\| + c_5(K,l) \|K'-K\| \\
	&\leq L_2 \|X'-X\|
	\end{aligned}
	\end{equation}
\end{proof}
Setting $\gamma_X = \min \{\gamma_{1},\gamma_{2} \}$ and $\zeta_X = L1, \beta_X = L2$, Lemma \ref{lem:lip} is proved.

\section{Proof of Theorem \ref{theorem:duality}}	

The following lemma from the duality theory~\citep{nesterov2013introductory,nedic2009subgradient} provides a sufficient and necessary condition for the absence of duality gap.

\begin{lemma}\label{lem:saddle}
	Suppose that $(X^*, \mu^*)$ is a feasible pair of the Lagrangian function $\mathcal{L}(X,\mu)$ with $X^* \in \mathcal{S}$ and $\mu^* > 0$, then the following three statements are equivalent:
	
	(a) $(X^*, \mu^*)$ is a saddle point for the Lagrangian function $\mathcal{L}(X,\mu)$.
	
	(b) $X^*$ and $\mu^*$ are optimal solutions to the primal and dual problems, respectively, with zero duality gap, i.e., $D^* = P^*$.
	
	(c) The following conditions hold, i.e.,
	\begin{equation}\label{equ:saddle}
	\begin{aligned}
	&\mathcal{L}(X^*,\mu^*) = \min_{X \in \mathcal{S}}\mathcal{L}(X,\mu^*), \\
	&J_c(X^*) \leq \bar{\rho}, \\
	&\mu^*(J_c(X^*)- \bar{\rho})=0.
	\end{aligned}
	\end{equation}
\end{lemma}

In the sequel, we prove Theorem \ref{theorem:duality} by examining the conditions in (\ref{equ:saddle}).

Define 
\begin{equation}\label{def:mu}
\mu^{*} \triangleq \inf \left\{\mu \geq 0: J_c\left(X^*(\mu)\right) \leq \bar{\rho}\right\}.
\end{equation}
We will show that when $\mu^*$ is finite, then the policy-multiplier pair $(X^*(\mu^*), \mu^*)$ satisfies the conditions in (\ref{equ:saddle}). And it is indeed the case if the Slater's condition holds.

Suppose that the Slater's condition is satisfied, i.e., there exists a feasible policy $X' \in \mathcal{S}$ such that $J_c(X') < \bar{\rho}$. We first show $\mu^*$ defined in (\ref{def:mu}) is finite.

Clearly, $J(X') < \infty$ since $X' \in \mathcal{S}$. By the definition of $D(\mu)$, for all $\mu \geq 0$ we have 
$$
D(\mu) \leq J(X') + \mu (J_c(X') - \bar{\rho}).
$$
Suppose that for any $\mu \geq 0$, $J_c(X^*(\mu))>\bar{\rho}$. Then, it follows that
\begin{equation}\notag
\begin{aligned}
J(X') &\geq \sup_{\mu \geq 0} D(\mu) - \mu (J_c(X') - \bar{\rho}) \\
&= \sup_{\mu \geq 0} J(X^*(\mu)) + \mu (J_c(X^*(\mu)) - \bar{\rho}) - \mu (J_c(X') - \bar{\rho})\\
&= \sup_{\mu \geq 0} J(X^*(\mu)) + \mu (J_c(X^*(\mu)) - J_c(X')) \\
&= \infty,
\end{aligned}
\end{equation}
which contradicts with $J(X') < \infty$. Thus, $\mu^*$ must be finite.

To show that the complementary slackness $\mu^*(J_c(X^*(\mu^*))- \bar{\rho})=0$ is satisfied, we discuss two cases. If $\mu^* = 0$, then the complementary slackness trivially holds; or $\mu^* > 0$, then we must have $J_c(X^*(\mu^*))=0$. Thus, it suffices to consider the second case, i.e., $\mu^* > 0$. By assumption, it follows that $J_c(X^*(0)) \geq \bar{\rho}$. Since $J_c(X^*(\mu))$ is decreasing with $\mu>0$, there exists a multiplier $\mu'$ such that $J_c(X^*(\mu))<\bar{\rho}$. We notice that $X^*(\mu)$ is continuous with respect to $\mu$ as all matrix inverses in (\ref{equ:stationary}) are continuous. Combining with the smoothness of $J_c(X)$, it follows that we can only have $J_c(X^*(\mu^*))=\bar{\rho}$.

Now, the conditions in (\ref{equ:saddle}) hold and there is zero duality gap.

\section{Convergence Analysis}
In this section, we provide the convergence analysis of the proposed algorithms.
\subsection{Proof of Theorem \ref{theorem:learning}}
We prove it by applying a result in the zero-order optimization~\cite[Theorem 1]{malik2019derivative}. To guarantee the convergence of random search, it requires (a) the gradient dominance, (b) the locally Lipschitz property and (c) the boundedness of gradient norm $G_{\infty} ~\text{and}~ G_{2}$. To this end, we only need to establish (c) by leveraging the uniform boundedness of $\|w_t\|$. Since $G_{2} \leq G_{\infty}^{2}$, it suffices to bound $G_{\infty}$.

We first show that $\widehat{\mathcal{L}}(X, \mu)$ is bounded for $X \in \mathcal{S}_{10}$. By the linear dynamics (\ref{equ:sys}), for $X \in \mathcal{S}$ the state $x_t$ can be written as a function of $x_0$ and $\{w_0, w_1, \dots\}$, i.e.,
$$
x_t = (A-BK)^tx_0 + \sum_{k = 0}^{t-1}(A-BK)^{t-1-k}(Bl+w_t).
$$

For a bounded noise sequence $w=\{w_0, w_1, \dots\}$, we have
\begin{align*}
\widehat{\mathcal{L}}(X, \mu) &= \lim\limits_{T \rightarrow \infty} \frac{1}{T} \sum_{t = 0}^{T}x_t^{\top}(Q_{\mu}+K^{\top}RK)x_t \\
&~~~~~~~~~~~~ + 2(S-2K^{\top}Rl)^{\top}x_t + l^{\top}Rl\\
&\leq \max \limits_{w} \lim\limits_{t \rightarrow \infty}x_t^{\top}(Q_{\mu}+K^{\top}RK)x_t \\
&~~~~~~~~~~+ 2(S-2K^{\top}Rl)^{\top}x_t + l^{\top}Rl
\end{align*}

Noting that $\lim\limits_{t \rightarrow \infty} (A-BK)^t x_0 = 0$, we have
\begin{align*}
&\widehat{\mathcal{L}}(X, \mu) \leq \max \limits_{w}(\sum_{k=0}^{\infty}(A-BK)^k(Bl+w_t))^\top\\
& \times (Q_\mu + K^{\top}RK)(\sum_{k=0}^{\infty}(A-BK)^k(Bl+w_t)) \\
&+ 2(\sum_{k=0}^{\infty}(A-BK)^k(Bl+w_t))^\top (S-K^{\top}Rl) + l^{\top}Rl \\
&\leq (\sum_{k=0}^{\infty}\|A-BK\|^k)^2(\|Bl\|+\|v\|)^2\|Q_\mu + K^{\top}RK\|\\
&+ 2(\sum_{k=0}^{\infty}\|A-BK\|^k)(\|Bl\|+\|v\|) \|S-K^{\top}Rl\| + l^{\top}Rl \\
&\leq \frac{1}{(1-\rho(A-BK))^2}(\|Bl\|+\|v\|)^2\|P_K\|+ l^{\top}Rl\\
&+ \frac{2}{1-\rho(A-BK)}(\|Bl\|+\|v\|) (\|S\|-\|K\|\|R\|\|l\|),
\end{align*}
where $\|P_K\|, \|K\|, \|l\|,\rho(A-BK)$ are uniformly bounded over $X \in \mathcal{S}$. Thus, $\widehat{\mathcal{L}}(X, \mu)$ is bounded.

Similarly, it follows from the Lipschitz property that
$$
\| \widehat{\mathcal{L}}(X+rU,\mu) - \mathcal{L}(X+rU,\mu) \| \leq F
$$
with $F>0$.

For a given radius $r < \gamma_0$ and a unit perturbation $U \in \mathbb{S}$, the gradient estimate is bounded as
\begin{align*}
\|\widehat{\nabla \mathcal{L}}\|_{2} &= \frac{n}{r^2}\|\widehat{\mathcal{L}}(X +rU, \mu)\|\\
&= \frac{n}{r^2}(\|\mathcal{L}(X +rU, \mu)\| + F)\\
& \leq \frac{n}{r^2} (\|\mathcal{L}(X, \mu)\| + r\xi_0 +F)\\
& \leq \frac{n}{r^2} (\|10\mathcal{L}(X_0, \mu)\| + \gamma_0\xi_0 + F).
\end{align*}

Now, the proof follows directly from \cite[Theorem 1]{malik2019derivative}.
\subsection{Proof of Theorem \ref{theorem:primal-dual}}

By the definition of projection and the subgradient, it follows that
\begin{align*}
\mathbb{E}[\|\mu_{i+1} - \mu^*\|^2] &\leq \mathbb{E}[\| \mu_i - \mu^* + \xi_i \cdot \hat{\omega}_i\|^2] \\
&= \mathbb{E}[\|\mu_i - \mu^*\|^2 +2\xi_i \hat{\omega}_i(\mu_i - \mu^*) + (\xi_i)^2\|\hat{\omega}_i\|^2]\\
&\leq \mathbb{E}[\|\mu_i - \mu^*\|^2] +2\xi_i\mathbb{E}[D(\mu_i) - D^*] + (\xi_i)^2b^2.
\end{align*}

Then, rearranging it yields that
\begin{align*}
\mathbb{E}[D^*-D(\mu_i)] \leq \frac{\mathbb{E}[\|\mu_i - \mu^*\|^2]}{2\xi_i} -\frac{\mathbb{E}[\|\mu_{i+1} - \mu^*\|^2]}{2\xi_i} + \frac{\xi_i b^2}{2}.
\end{align*}

Summing up from $i=1$ to $k$ and noting $\xi_i \geq \xi^{i+1}$, it follows that
\begin{align*}
&\mathbb{E}[\sum_{i=1}^{k}(D^*-D(\mu_i))] \leq -\frac{1}{2\xi_{j+1}}\mathbb{E}[\|\mu_{j+1}-\mu^*\|] + \frac{b^2}{2} \sum_{i=1}^{j} \xi_i \\
&~~~+ \frac{1}{{2\xi_1}}\mathbb{E}[\|\mu_{1} - \mu^*\|^2] +  \frac{1}{2}\sum_{i=1}^{j}(\frac{1}{\xi_{i+1}} - \frac{1}{\xi_i})\mathbb{E}[\|\mu_{i+1}-\mu^*\|^2] \\
&\leq \frac{2}{\xi^j}e^2 + \frac{b^2}{2} \sum_{i=1}^{j} \xi_i.
\end{align*}

By Jenson's inequality, one can easily obtain that
\begin{align*}
\mathbb{E}[D^*-	D(\bar{\mu}_j)] \leq \frac{2}{j\xi_j}e^2 + \frac{b^2}{2j} \sum_{i=1}^{j} \xi_i.
\end{align*}

The proof follows by noting that $\xi_i = \frac{1}{be} \sqrt \frac{2}{i}$.

\end{appendices}
\end{document}